\documentclass[prd,twocolumn,nofootinbib,showpacs,eqsecnum,floatfix,superscriptaddress]{revtex4}

\usepackage{amsmath}
\usepackage{graphicx}
\usepackage{bbm}
\usepackage{color}

\newcommand{\eq}[1]{\begin{equation}\label{#1}}
\newcommand{\en}{\end{equation}}
\newcommand{\1}{\mathbbm{1}}

\newcommand{\SU}{\mathrm{SU}}
\newcommand{\dks}{D_\text{KS}}
\DeclareMathOperator{\tr}{tr}

\begin{document}

\title{Polyakov loops and spectral properties of the staggered Dirac operator}

\author{Falk Bruckmann} 
\affiliation{Institute for Theoretical Physics, University of
Regensburg, 93040 Regensburg, Germany}

\author{Stefan Keppeler} 
\affiliation{Institute for Theoretical Physics, University of
Regensburg, 93040 Regensburg, Germany} 
\affiliation{Mathematics Institute, University of T\"ubingen, Auf der
Morgenstelle 10, 72076 T\"ubingen, Germany}

\author{Marco Panero}  
\affiliation{Institute for Theoretical Physics, University of
Regensburg, 93040 Regensburg, Germany}

\author{Tilo Wettig}  
\affiliation{Institute for Theoretical Physics, University of
Regensburg, 93040 Regensburg, Germany}

\date{15 August 2008}

\begin{abstract}
  We study the spectrum of the staggered Dirac operator in SU(2) gauge
  fields close to the free limit, for both the fundamental and the
  adjoint representation. Numerically we find a characteristic cluster
  structure with spacings of adjacent levels separating into three
  scales. We derive an analytical formula which explains the emergence
  of these different spectral scales. The behavior on the two coarser
  scales is determined by the lattice geometry and the Polyakov loops,
  respectively. Furthermore, we analyze the spectral statistics on all
  three scales, comparing to predictions from random matrix theory.
\end{abstract}



\maketitle 

\section{Introduction}
\label{introsect}

One of the goals of quantum chromodynamics (QCD), the fundamental
theory of quarks and gluons, is the calculation of the hadronic mass
spectrum from first principles. For this purpose, a nonperturbative
regularization of QCD can be formulated on a space-time lattice, which
makes the theory mathematically well defined and accessible for
numerical simulations. In order to describe fermionic fields, one needs
a discretized Dirac operator, which can be constructed in different
ways.  In particular, the staggered or Kogut-Susskind Dirac operator
\cite{Kogut:1974ag} is widely used as it is computationally cheaper
than other options.  (We shall not enter the debate of the rooting
issue \cite{Creutz:2007rk,Kronfeld:2007ek} here.)

During the past 15 years it has been shown that some universal
properties of the QCD Dirac spectrum can be described by a version of
random matrix theory (RMT) \cite{MehtaBook,GMW97} which incorporates
the chiral symmetry of the massless Dirac operator and is accordingly
called chiral random matrix theory (chRMT)
\cite{Shuryak:1992pi,Verbaarschot:2000dy}. In chRMT, one models the
Dirac operator by a block off-diagonal matrix with random entries,
respecting the global symmetries of the massless Dirac operator. The
path integral over the gauge fields is then replaced by averaging over
an ensemble of such matrices. Chiral RMT reproduces low-energy sum
rules \cite{Verbaarschot:1994gr} and yields accurate predictions for,
e.g., the microscopic spectral density \cite{Verbaarschot:1993pm}, the
distribution of the smallest eigenvalues
\cite{Nishigaki:1998is,Damgaard:2000ah}, level spacing distributions,
and other short-range spectral correlations.  Predictions from chRMT
have been successfully compared with numerical results from lattice
gauge theory in many different settings
\cite{Halasz:1995vd,Verbaarschot:1995yi,BerbenniBitsch:1997tx,BerbenniBitsch:1998sy,Damgaard:1998ie,Edwards:1999px,Edwards:1999ra,Shcheredin:2003um,Giusti:2003gf,Fukaya:2007fb}.

When using RMT to predict spectral properties of complex quantum
systems, it is essential that the random matrices have the same
antiunitary symmetries as the system to be modeled, see, e.g.,
Ref.~\cite{HaakeBook}. One distinguishes different symmetry classes
and corresponding ensembles of random matrices, labeled by their Dyson
index $\beta_D=1$, $2$, or $4$ \cite{MehtaBook}.

The staggered Dirac operator on the lattice exhibits the peculiar
feature that its symmetry properties can be different from that of the
continuum theory. In particular, for gauge group $\SU(2)$ with
fermions in the fundamental representation and for gauge group
$\SU(N)$ with fermions in the adjoint representation, the $\beta_D=1$
and $\beta_D=4$ cases are interchanged compared to the continuum Dirac
operator (see below). This implies that the spectral properties of the
staggered operator are different from those of the continuum Dirac
operator. Accordingly, a transition is expected to take place in the
continuum limit. The first indication of such a transition has been
reported in Ref.~\cite{Follana:2006zz}.

In this work, we study a related but different case, namely, the
staggered Dirac operator for fermions in the fundamental and adjoint
representation of $\SU(2)$ in the free limit. This limit is approached
by increasing the Wilson gauge action parameter $\beta=4/g^2$ at fixed
(or mildly varying) lattice size, i.e., the lattice spacing, and thus
the physical volume, shrinks to zero. 

In the free limit the dynamics becomes integrable and therefore one no longer expects RMT statistics. Consequently, in this limit we do not expect the same transition between RMT symmetry classes as in the continuum limit described above. Instead, generically, the eigenvalues should be uncorrelated, like numbers drawn from a Poisson process
\cite{BerTab77,Bohigas:1983er}.  It turns out that in the case we are
studying, the situation is somewhat more complicated: Close to the
free limit it is possible to disentangle three different scales that
appear in the separations between the eigenvalues.

First, the eigenvalues form well-separated plateaux centered at the
eigenvalues of the free staggered operator.

Second, we observe an internal structure of the plateaux: The
eigenvalues arrange themselves in clusters of eight eigenvalues each.
We will show that, for each configuration, the location of these
clusters can be predicted from the knowledge of only four real
variables, i.e., the averaged traced Polyakov loops in the four
lattice directions.

Third, on the finest scale, the eigenvalue fluctuations within
clusters can be described in terms of chRMT. We will present numerical
results for the level spacing densities which agree with the chRMT
prediction for all values of the Wilson gauge action parameter
$\beta$.

This article is organized as follows.  In
Sec.~\ref{antiunitarysymmetriessect} we briefly introduce some basic
notions of random matrix theory and discuss the different antiunitary
symmetries which are relevant for the Dirac operator in the continuum
and for the staggered Dirac operator on the lattice. In
Sec.~\ref{polyakovsect} we derive an analytical prediction for the
staggered Dirac eigenvalues in certain gauge field configurations
close to the free limit.  Section~\ref{numericalsec} begins with a
comparison of numerical data from lattice simulations to our
analytical prediction and continues with an analysis of spectral
statistics on all three scales; i.e., we study the distribution of
level spacings within clusters, between clusters, and between
plateaux.  Wherever appropriate we compare to the predictions from
chRMT and from a Poisson process.  Throughout
Secs.~\ref{antiunitarysymmetriessect} to \ref{numericalsec} we discuss
the fundamental and adjoint representation side by side.  In Sec.~V we
conclude with a discussion of the implications of our findings.
Preliminary results of this study have been presented in
Ref.~\cite{Bruckmann:2008rj}.

\section{Antiunitary symmetries and RMT ensembles}
\label{antiunitarysymmetriessect}

The notion of universality, as commonly used in the context of RMT and the analysis of spectra of complex quantum systems,
means that spectral statistics can be described by an appropriate
ensemble of random matrices, which shares the symmetries of the system
under consideration. 
Depending on the presence of antiunitary symmetries, the entries of
matrices of the ensemble have to be either real, complex, or
quaternion real. The associated Dyson indices are $\beta_D=1$, $2$, or
$4$, respectively.

In particular, RMT yields a prediction for the universal quantity
$P(s)$, the probability density for the unfolded nearest-neighbor
spacings $s$ (see Sec.~\ref{sec:unfolding} for a discussion of
unfolding). This prediction is well approximated by the Wigner
surmise,
\eq{wignersurmise}
P(s) = a\,s^{\beta_D}e^{-bs^2} \;,
\en
where
\eq{aandb}
a = 2\,\frac{\Gamma^{\beta_D+1} \left( \beta_D/2 +1 \right)}
            {\Gamma^{\beta_D+2} \left( (\beta_D+1)/2 \right)}\, ,\; 
b = \frac{\Gamma^2\left( \beta_D/2+1 \right)}
         {\Gamma^2 \left( (\beta_D +1)/2 \right)}\,. 
\en
The Wigner surmise is the level spacing density for ensembles of
$2\times2$ matrices with Dyson index $\beta_D$.

For the QCD Dirac operator $D$ the RMT description is formulated in
terms of matrices which reflect the chiral, flavor, and antiunitary
symmetries of $D$ \cite{Verbaarschot:1994qf}. For gauge group $\SU(N)$
with $N \geq 3$ colors and fermions in the fundamental representation,
$D$ generically has complex elements and does not commute with any
antiunitary operator. Accordingly, universal spectral correlations
are described by the chiral unitary ensemble (chUE), labeled by the
Dyson index $\beta_D=2$.

However, the Dirac operator enjoys invariance with respect to an
antiunitary transformation if the fermions are either in the
fundamental representation of the gauge group $\SU(2)$, or in the
adjoint representation of $\SU(N)$ with $N\geq 2$ arbitrary. These two
cases are discussed in more detail in the following.

\subsection{Fundamental representation}
\label{antiunitaryfundsubsect}

In a nutshell, the antiunitary invariance of the Dirac operator with
$\SU(2)$ gauge fields and fermions in the fundamental representation,
i.e., QCD with two colors, is based on the fact that the generators
are $\tau_a/2$, and the Pauli matrices $\tau_a$ possess the following
complex conjugation property:
\eq{paulicomplexconjugation}
\tau_a^\ast = - \tau_2 \tau_a \tau_2 \; .
\en
The antiunitary symmetry operator, however, is realized in different
ways in the continuum and in the lattice formulation with staggered
fermions.

\subsubsection{Continuum}
\label{antiunitaryfundcontsubsubsect}

In the continuum, the anti-Hermitian massless Dirac operator is
defined as
\eq{dcont}
D = \gamma_\mu D_\mu 
  = \gamma_\mu \left( \partial_\mu + i  A_\mu^a T_a\right)
\en
with $T_a=\tau_a/2$ for the fundamental representation of $\SU(2)$. It
anticommutes with the chirality operator $\gamma_5$, and therefore
its nonzero eigenvalues, which are purely imaginary, come in complex
conjugate pairs.

Using Eq.~(\ref{paulicomplexconjugation}) one easily verifies that the
operator $D$ is invariant under an antiunitary symmetry,
\eq{continuumchgoe} 
[\mathcal{C} \gamma_5 \tau_2 K, D ] = 0
\; , 
\en 
where $\mathcal{C} = \gamma_2 \gamma_4$ is the charge conjugation
matrix, and $K$ denotes complex conjugation (in the position
representation). Note that $\mathcal{C} \gamma_5$ acts on the spinor
indices, whereas $\tau_2$ acts in color space. Since $( \mathcal{C}
\gamma_5 \tau_2 K )^2 = \1$, it follows that $D$ can be made real by a
basis transformation that does not depend on the gauge configuration
\cite{PorterBook,HaakeBook}. Accordingly, the RMT description for
two-color QCD and fundamental fermions is formulated in terms of the
chiral orthogonal ensemble (chOE), characterized by $\beta_D=1$.

\subsubsection{Lattice}
\label{antiunitaryfundlatsubsubsect}

The staggered or Kogut-Susskind Dirac operator for a hypercubic
lattice of finite spacing $a$ in $d$ dimensions is given by
\eq{dks}
\left( \dks \right)_{x,y} 
= \frac{1}{2a} \sum_{\mu=1}^{d} \eta_\mu(x) \left[ 
  \delta_{x+\hat\mu,y} U_\mu^\dagger (x) 
  - \delta_{x-\hat\mu,y} U_\mu(y) \right] 
\en
with $\eta_\mu(x)=(-1)^{\sum_{\nu<\mu} x_\nu}$ and $U_\mu(x) \in
\SU(2)$. On the lattice the operator
$S=\delta_{x,y}(-1)^{\sum_{\nu=1}^d x_\nu}$ plays the same role as
$\gamma_5$ does in the continuum: Since $\{ \dks, S \} = 0$, the
eigenvalues of $\dks$ also come in (purely imaginary) complex
conjugate pairs.

The staggered Dirac operator, which is widely used in numerical
simulations (because it maintains a remnant of chiral symmetry,
partially solves the doubling problem, and is computationally cheaper
than other lattice Dirac operators) exhibits the peculiar feature
that its antiunitary symmetries are different from those of the
continuum Dirac operator
\cite{KlubergStern:1982bs,Hands:1990wc,Halasz:1995vd}.  Because of Eq.~(\ref{paulicomplexconjugation}) the links obey $U_\mu(x)=\tau_2
U_\mu^{\ast}(x) \tau_2$. Since the $\gamma$ matrices have been replaced by real numbers, $\eta_\mu(x)$, no charge conjugation is required in order to compensate for the complex
conjugation. Therefore, $ \dks$ is invariant under the following
antiunitary symmetry:
\eq{tau2kdks} [ \tau_2 K, \dks ] = 0
\; .  
\en
As $( \tau_2 K )^2 = - \1 $, it follows that $\dks$ can always be
written as a quaternion real matrix \cite{HaakeBook}. Hence, the
chiral symplectic ensemble (chSE), with $\beta_D=4$, is used to
describe its universal properties. Another consequence of invariance
with respect to an antiunitary transformation with square $-\1$ is
Kramers' degeneracy, see, e.g., \cite{SakuraiBook,HaakeBook}; i.e.,
all eigenvalues have (at least) multiplicity two. This degeneracy has
to be removed by hand before one discusses spectral correlations.

\subsection{Adjoint representation}
\label{antiunitaryadjsubsect}

In this case, the antiunitary symmetries are determined by the purely
imaginary nature of the matrix elements of the generators in the
adjoint representation (being the structure constants) of the gauge
group $\SU(N)$.  As in the previous case, the way this symmetry is
realized in the continuum and for the staggered Dirac operator on the
lattice is different.

\subsubsection{Continuum}
\label{antiunitaryadjcontsubsubsect}

The purely imaginary adjoint generators induce a real covariant
derivative in the continuum Dirac operator of Eq.~(\ref{dcont}), which
is therefore invariant under the following antiunitary symmetry:
\eq{cgamma5kd}
[\mathcal{C} \gamma_5  K, D ] = 0 \; .
\en
Since $( \mathcal{C} \gamma_5 K )^2 = -\1$, it is possible to recast
$D$ into real quaternionic form. This implies that the Dirac spectrum
can be described in terms of the chiral symplectic ensemble (chSE),
labeled by $\beta_D=4$.

\subsubsection{Lattice}
\label{antiunitaryadjlatsubsubsect}

For fermions in the adjoint representation, the staggered Dirac
operator $\dks$ is explicitly real, because such are the $U_\mu(x)$
link matrices appearing on the right-hand side of Eq.~(\ref{dks}),
which now take values in the adjoint representation of $\SU(2)$. This
implies that the appropriate RMT ensemble is the chOE, characterized
by the Dyson index $\beta_D=1$.

\section{Spectral predictions from Polyakov loops}
\label{polyakovsect}

The spectral properties of the Dirac operator are relevant to the two
major unsolved nonperturbative problems in QCD: confinement and
chiral symmetry breaking. On the one hand, the eigenvalue density is
related to the chiral condensate, the order parameter for the chiral
phase transition, by the Banks-Casher relation \cite{Banks:1979yr}. On
the other hand, the average value of the Polyakov loop, which, in the
quenched case, is an order parameter for the confinement-deconfinement
transition, can be expressed through sums of the eigenvalues of the
lattice Dirac operator with different boundary conditions
\cite{Gattringer:2006ci,Bruckmann:2006kx,Synatschke:2007bz,Bilgici:2008qy}.

On commensurate lattices, i.e., on lattices for which the numbers
$L_\mu/2$, $\mu=1,\hdots,d$, are rationally dependent (the staggered
Dirac operator only makes sense on lattices with even $L_\mu$), the
spectrum of $\dks$ in the trivial vacuum, i.e., for the field
configuration with all $U_\mu(x)$ equal to unity, is highly
degenerate. Our analysis of the spectrum close to the free limit shows
that the way in which this degeneracy is (partially) lifted can be
expressed in terms of the traced and averaged Polyakov loops in all
directions,
\eq{dummy}
P_\mu = \frac{1}{2} \, \tr \left\langle 
         \prod_{n=1}^{L_\mu} U(x + n \hat{\mu}) \right\rangle_x \, , 
\en 
where the $x$ average is over the whole lattice.  For each
configuration, this effect can be predicted by calculating the
spectrum obtained from a vacuum configuration with uniform link
variables in each direction, taking values in an Abelian subgroup of
the gauge group, and yielding the same averaged Polyakov loops as the
original configuration.

\subsection{Fundamental representation}
\label{polyakovfundsubsect}

For the trivial gauge vacuum (all links set to $\1$ or any gauge
transform thereof), the eigenvalues of the staggered Dirac operator
read
\eq{trivialeigenvalues}
\lambda_n^\pm 
  = \pm i \sqrt{\sum_{\mu=1}^d  \sin^2 \left[ 
            \frac{2 \pi}{L_\mu} \left( k_\mu + c_\mu \right) \right] } \, , 
\en
with the wave numbers taking integer values $ 0 \le k_\mu < L_\mu/2$
and $c_\mu = 0$ $\left(c_\mu= \frac{1}{2}\right)$ for (anti)periodic
boundary conditions for the fermionic wave function in direction
$\mu$.

However, the system also admits vacua in different center sectors,
which can be labeled by the traced Polyakov loops $P_\mu = \pm 1$ [for
$\SU(2)$]. The sign of the latter can always be inverted by a
multiplication of all the links in the $\mu$ direction in a given
fixed-$x_\mu$ slab by $-\1$. In the Dirac operator this can be
compensated for by switching from periodic to antiperiodic boundary
conditions (or vice versa) in the direction $\mu$. Therefore, when we
consider a vacuum where the Polyakov loop in direction $\mu$ is $-1$,
we can equivalently set $c_\mu = \frac{1}{2}$ $\left(c_\mu=0\right)$
for (anti)periodic boundary conditions of the Dirac operator
in direction $\mu$.

For a generic lattice in $d$ dimensions, the number of possible free
spectra is thus equal to the number of allowed topological sectors for
the vacuum, i.e., $2^d$. When at least two lattice extensions are
equal, the number of possible free spectra is reduced. For lattices
with $L_\mu = L_\nu$ $\forall\,\mu,\nu=1,\hdots, d$, the different
patterns are labeled by the number of $c_\mu$ values that are equal to
$\frac{1}{2}$, thus yielding $d+1$ inequivalent possibilities.

Configurations close to the free limit are expected to approach
(modulo gauge transformations) one of the possible free vacua. This is
confirmed by our lattice simulations, where we find the distribution
of $P_\mu$ to be peaked at $\pm 1$. The corresponding free vacua can
then easily be identified by the sign of the averaged traced Polyakov
loops $P_\mu$ in the various directions.  Accordingly,
Eq.~(\ref{trivialeigenvalues}) provides a first approximation to the
observed spectrum of the staggered Dirac operator.

This prediction can be refined as follows. For a given configuration,
let us introduce a configuration built from uniform links $U_\mu(x)
\equiv U_\mu$ in each direction, taking values in an Abelian subgroup
of $\SU(2)$ (for instance the diagonal one), and yielding the same
averaged traced Polyakov loops as the original configuration. Since
for uniform and commuting links all plaquettes are equal to unity
these configurations may also be called vacuum configurations.

For these vacuum configurations, the gauge transformation
\eq{gaugetransform}
\begin{split} 
&U_\mu(x) \mapsto g(x) \, U_\mu(x) \, g^\dag(x+\hat{\mu})\\
&\text{with} \quad 
g(x)=\prod_{\mu=1}^d (U_\mu)^{x_\mu}
\end{split} 
\en
can be used in order to trivialize all Polyakov loops to $P_\mu=1$ at
the expense of introducing periodicity only up to $g(x)$ (with
$x_\mu=L_\mu$).  The latter equals the original Polyakov loop $P_\mu$
and (in the diagonal subgroup) comprises two opposite phases, which
behave like the constants $c_\mu$ (this is a generalization of our
previous argument that $P_\mu=-1$ can be absorbed by switching $c_\mu$
between $0$ and $\frac{1}{2}$). The spectrum of the staggered Dirac
operator in such a vacuum configuration is given by
\eq{clustersandpolyakovloops}
\lambda_n^\pm 
  = \pm i \sqrt{\sum_{\mu=1}^d \sin^2 \left[ \frac{2 \pi}{L_\mu}
      \left( k_\mu + c_\mu + \frac{\arccos P_\mu}{2 \pi} \right)\right] }\:.
\en
Now $c_\mu$ is again fixed to $0$ ($\frac{1}{2}$) for (anti)periodic
boundary conditions in the direction $\mu$. Our expectation is that,
close to the free limit, Eq.~(\ref{clustersandpolyakovloops}) provides
a better approximation than Eq.~(\ref{trivialeigenvalues}) to the
observed spectrum of $\dks$.

The eigenvalues in Eq.~\eqref{clustersandpolyakovloops} have a
multiplicity of $2^d$, which we derive in the following paragraph. For
dimension $d=4$, this implies an eightfold degeneracy in addition to
Kramers' degeneracy. A small perturbation of the vacuum generically
lifts this eightfold degeneracy but not Kramers' degeneracy, which
remains exact. This mechanism gives rise to what we will call clusters
of eight eigenvalues below.

In order to explain how the multiplicity of $2^d$ comes about, we have
to digress to a sketch of the derivation of
Eq.~\eqref{clustersandpolyakovloops}. The spectrum is most easily
obtained by looking at the square $\dks^2$ of the staggered Dirac
operator and applying it to plane waves. After some algebra, one finds
the eigenvalues of $\dks^2$ to be given by
\eq{spectrumofdks^2}
\Lambda_n 
  = - \sum_{\mu=1}^d \sin^2 \left[ \frac{2 \pi}{L_\mu}
      \left( k_\mu + c_\mu + \frac{\arccos P_\mu}{2 \pi} \right)\right] 
\en
with $0 \leq k_\mu \leq L_\mu-1$. Since $\Lambda_n$ is invariant under
$k_\mu \mapsto k_\mu + L_\mu/2$, we can restrict the wave numbers to
$0 \leq k_\mu \leq L_\mu/2-1$ and assign a multiplicity of $2^d$ to
each eigenvalue. An additional multiplicity factor of 2 arises from
the color degeneracy, yielding an overall multiplicity of $2^{d+1}$.
Because of the symmetry of the spectrum of $\dks$ about $\lambda = 0$
(which arose due to $\{\dks,S\}=0$, see
Sec.~\ref{antiunitaryfundlatsubsubsect} above), the eigenvalues of
$\dks$ are given by the positive and negative square roots of
$\Lambda_n$ with each eigenvalue $\lambda_n^\pm$ having half the
multiplicity of the corresponding $\Lambda_n$, i.e., $2^d$.

\subsection{Adjoint representation}
\label{polyakovadjsubsect}

The situation is similar for the adjoint representation of $\SU(2)$.
Since the latter is insensitive to the group center, all trivial vacua
are equivalent to the configuration with all links equal to unity.

The construction above can be repeated by considering a vacuum
configuration built from link matrices of the form
\eq{adjointslabrotation}
U_{\mu} = \left( 
\begin{array}{ccc}
\cos \alpha_\mu & - \sin \alpha_\mu & 0 \\
\sin \alpha_\mu & \cos \alpha_\mu & 0 \\
0 & 0 & 1 \\
\end{array}
\right) \; ,
\en
where $\alpha_\mu$ is related to the traced Polyakov loop $P_\mu$ by
$L_\mu\alpha_\mu=\arccos((P_\mu-1)/2)$. The analog of
Eq.~(\ref{clustersandpolyakovloops}) now reads
\eq{adjointclustersandpolyakovloops}
\lambda_n^\pm
  = \pm i \sqrt{\sum_{\mu=1}^d \sin^2 \left[ \frac{2 \pi}{L_\mu} 
          \left( k_\mu + c_\mu \right) + n  \alpha_\mu\right] } \;,
\en 
where $n \in \{ -1, 0, 1\}$ and $0 \leq k_\mu < L_\mu/2$. In contrast
to the situation for the fundamental representation,
Eq.~(\ref{adjointclustersandpolyakovloops}) predicts that one-third of
the eigenvalues remains unchanged, i.e., they are identical to the
eigenvalues in the trivial vacuum (the configuration with all links
equal to unity).

An analysis analogous to that for the fundamental representation shows
that now the multiplicity is $2^{d-1}$ (as opposed to $2^d$ above).
However, here we have no Kramers degeneracy, and thus a small
perturbation of the vacuum again gives rise to clusters of eight
eigenvalues in dimension $d=4$.

\section{Numerical results}
\label{numericalsec}

\begin{figure*}
  \includegraphics[width=.32\textwidth]{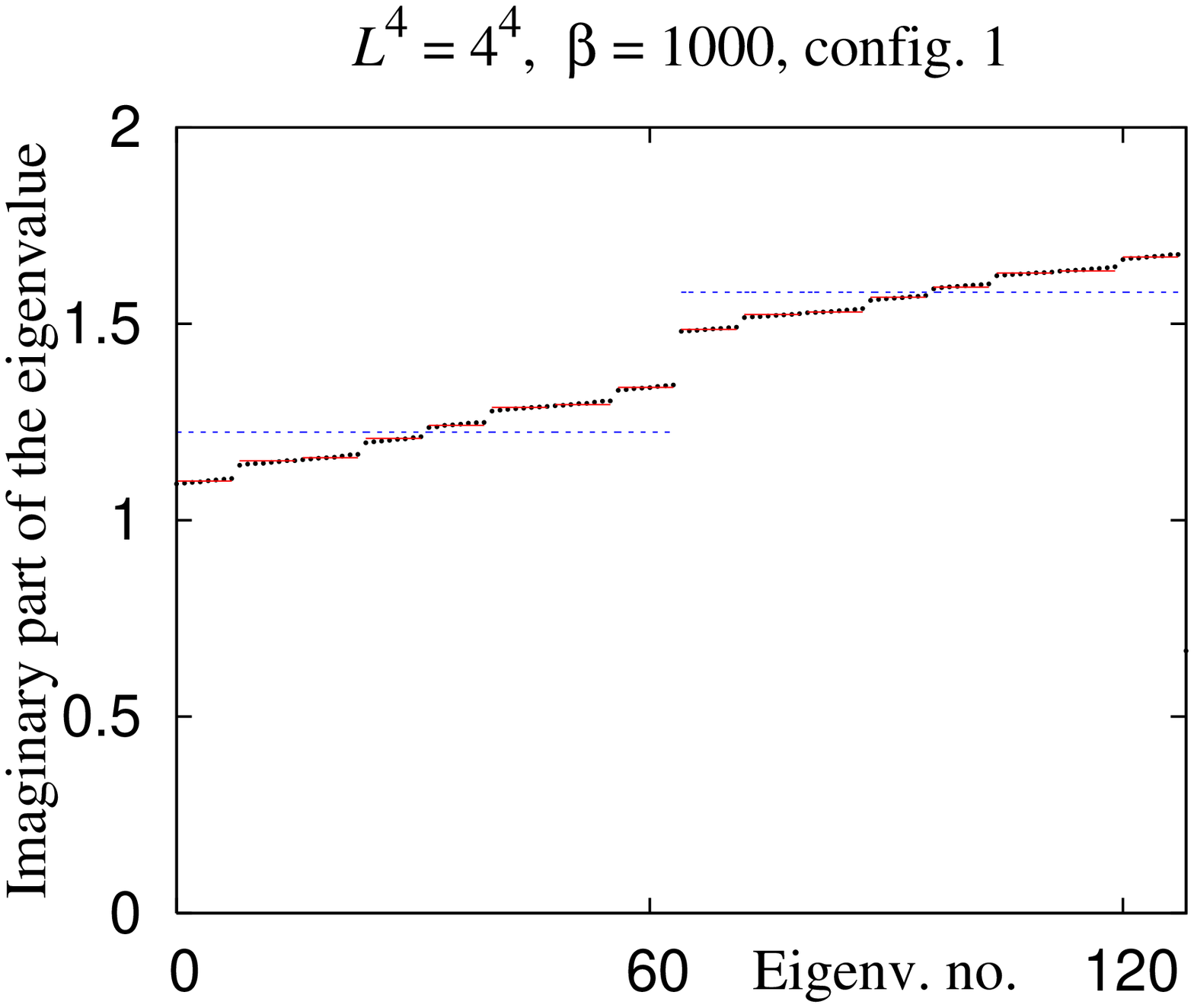}\hfill
  \includegraphics[width=.32\textwidth]{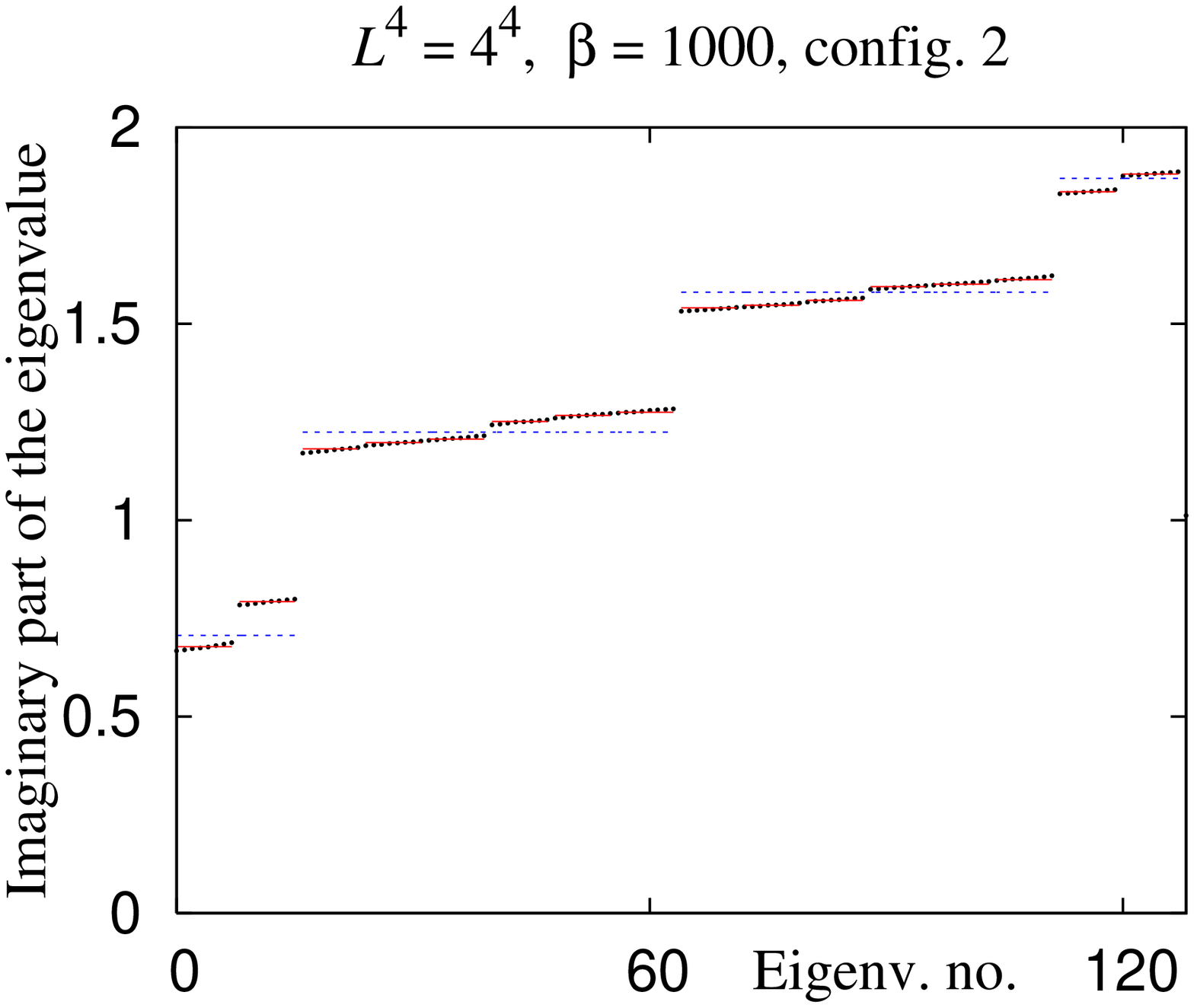}\hfill
  \includegraphics[width=.32\textwidth]{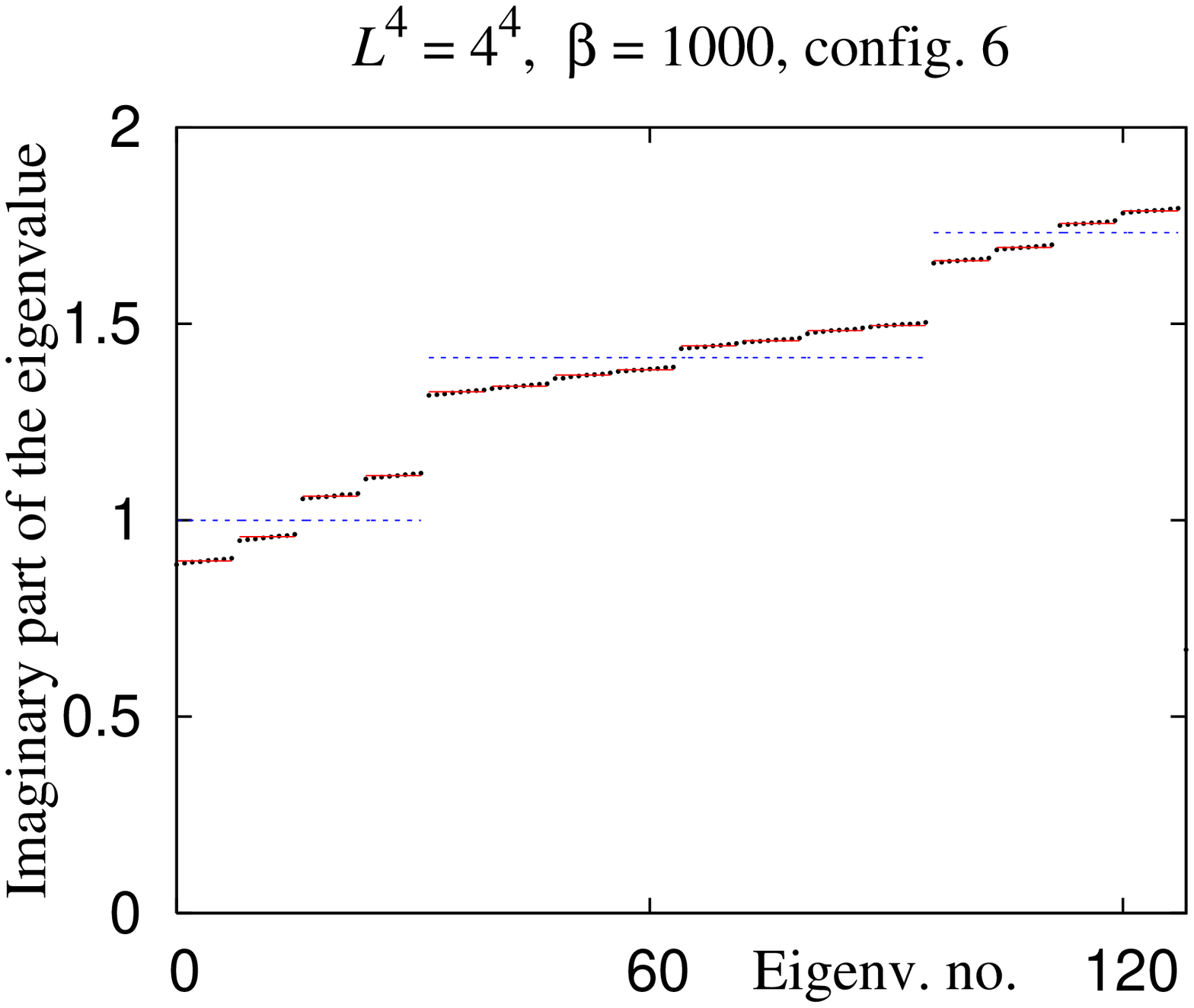}\\
  \includegraphics[width=.32\textwidth]{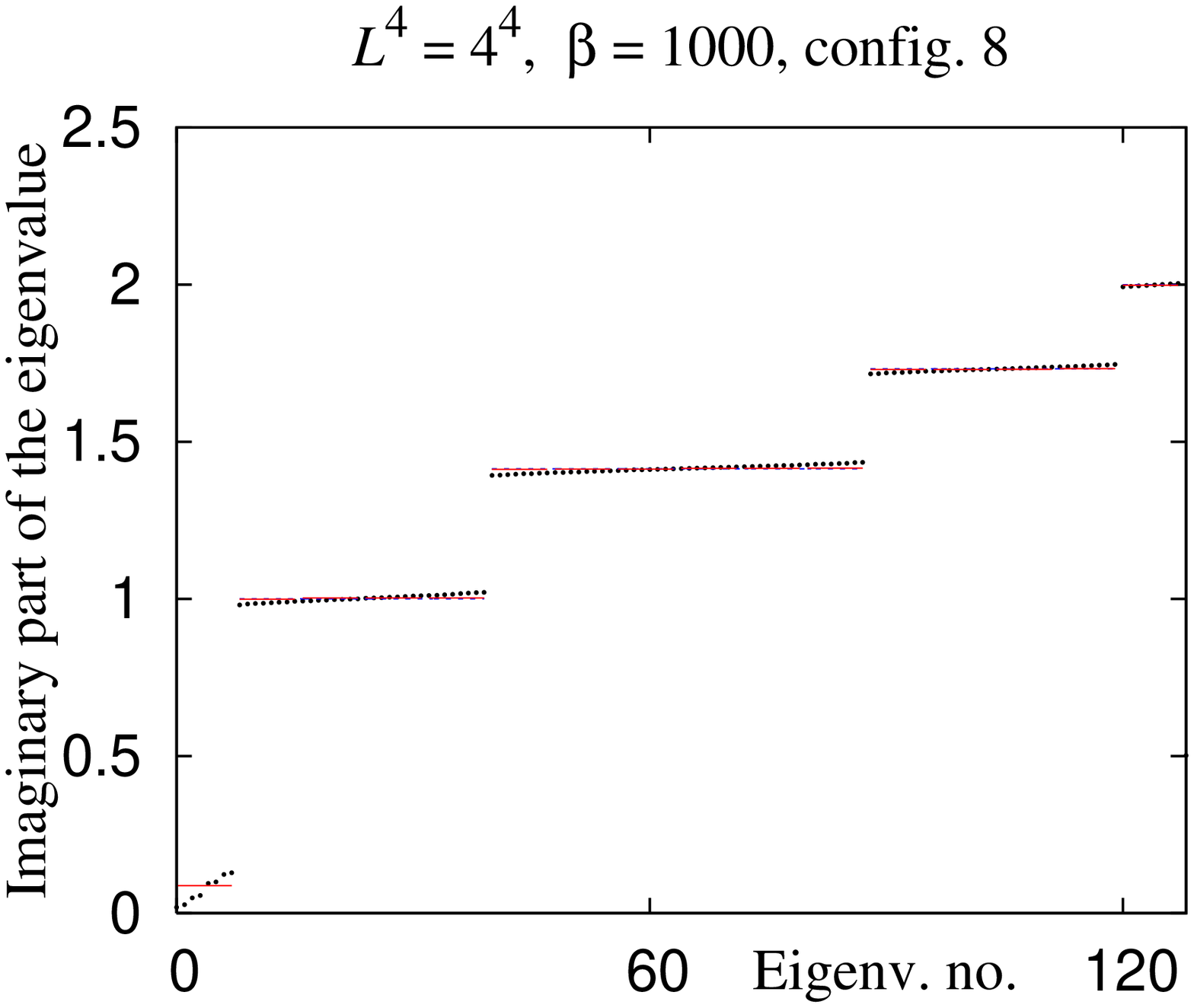}\hfill
  \includegraphics[width=.32\textwidth]{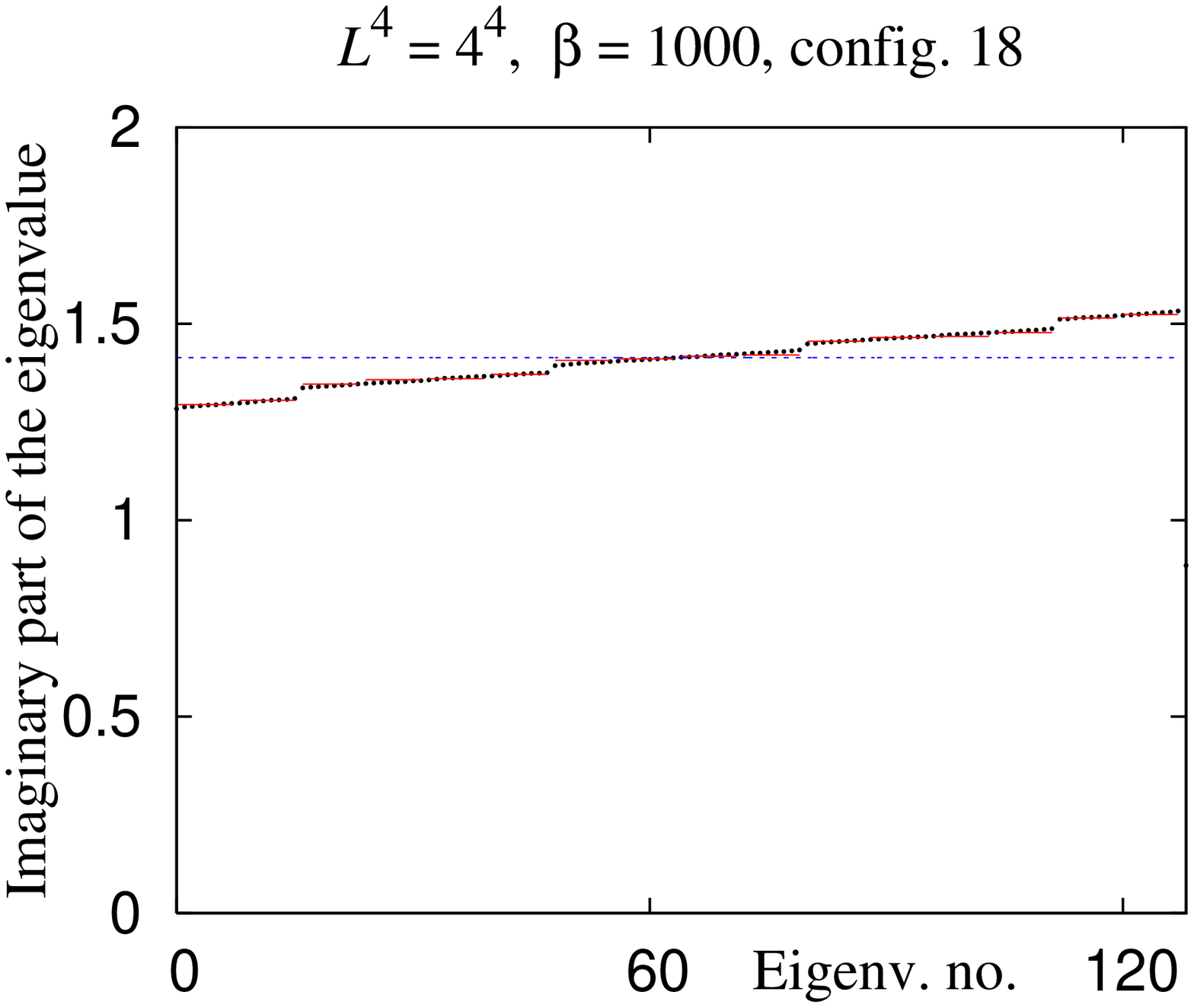}\hfill
  \includegraphics[width=.32\textwidth]{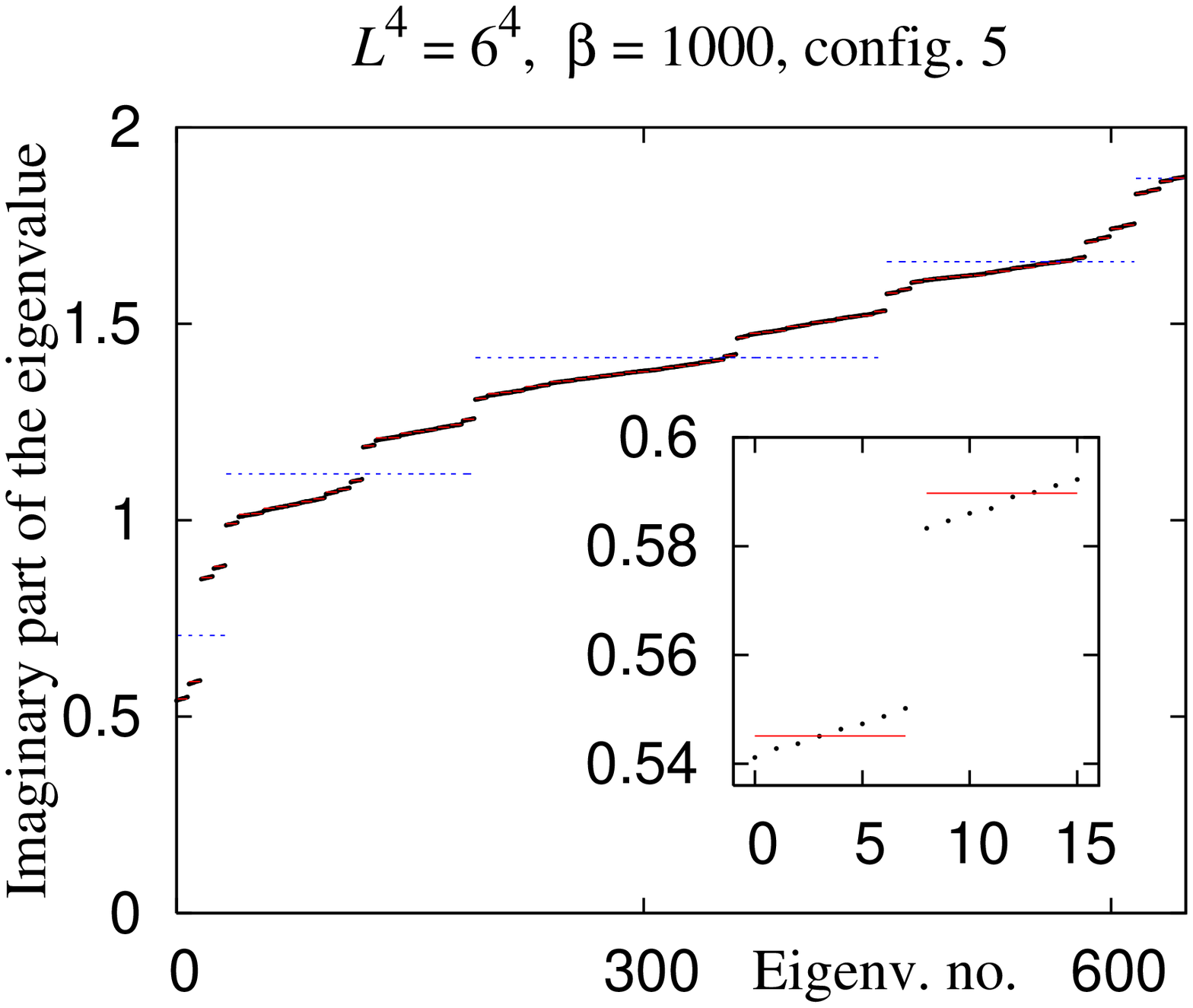}
  \caption{(Color online) Eigenvalues of $\dks$ close to the free limit for the
    fundamental representation of $\SU(2)$ ($L^4=4^4,\,\beta=1000$).
    We display representatives for the five different plateau patterns
    (indicated by dashed blue lines) described below
    Eq.~\eqref{trivialeigenvalues}. Within the plateaux the
    eigenvalues (black dots) arrange themselves in clusters of eight,
    whose locations are predicted by
    Eq.~(\protect\ref{clustersandpolyakovloops}) (solid red lines).
    The lower right panel shows a spectrum from a $6^4$ lattice, with
    a richer cluster structure.}
  \label{L4polyakovfig}
\end{figure*}

\subsection{Simulation details}

Our numerical results are obtained from sets of quenched $\SU(2)$
configurations generated using the Wilson gauge action. The simulation
algorithm is based on a combination of Metropolis and over-relaxation;
center rotations to explore different topological sectors are
implemented as well.

We obtain the full spectrum of the staggered Dirac operator from
ensembles of configurations on hyper-cubic, isotropic lattices with
volumes ranging from $V=4^4$ to $16^4$. The spectrum is evaluated
using the Cullum-Willoughby version of Lanczos' algorithm
\cite{CullumWilloughby}; periodic (antiperiodic) boundary conditions
are assumed in the spatial (temporal) directions.

For each lattice volume $V$ and for each value of the Wilson gauge
action parameter $\beta$, our analysis of the full staggered spectrum
is based on a number of thermalized and uncorrelated configurations
between a few tens and a few thousands. For each configuration the
number of distinct eigenvalues with positive imaginary part is $V/2$
for the fundamental representation and $3V/2$ for the adjoint
representation. Hence, our data for each pair $(V,\beta)$ typically
contains on the order of one million distinct eigenvalues.

Furthermore, we also generate gauge configurations on much larger
lattices (up to $ 34 \times 38 \times 46 \times 58$) in order to
investigate the distribution of eigenvalues from Eqs.
(\ref{clustersandpolyakovloops}) and
(\ref{adjointclustersandpolyakovloops}). On these lattice we do not
diagonalize the Dirac operator but calculate only the averaged
Polyakov loops.

\subsection{Separation of spectral scales}
\label{separationofscalessec}

\begin{figure*}
  \includegraphics[width=.32\textwidth]{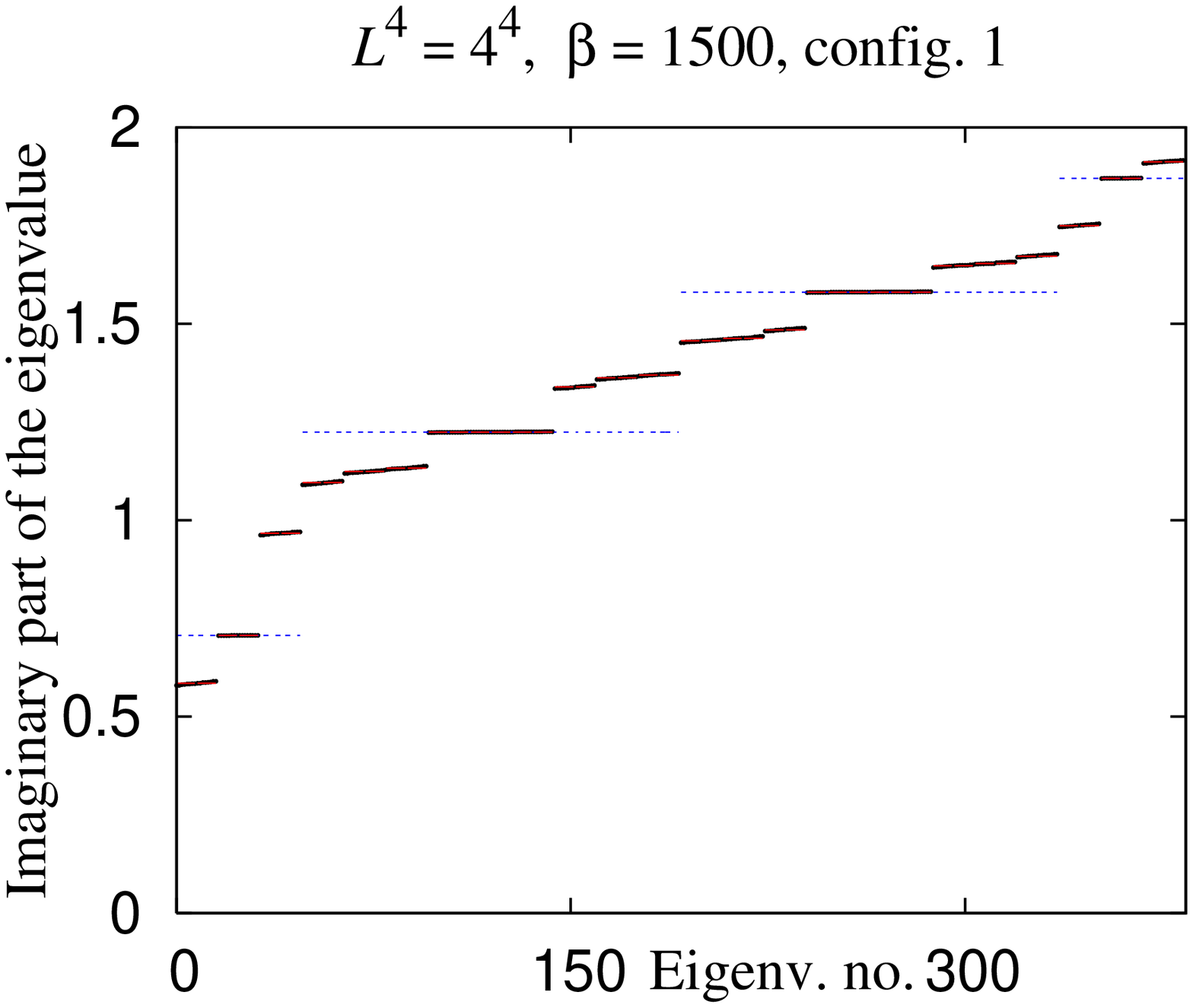}
  \includegraphics[width=.32\textwidth]{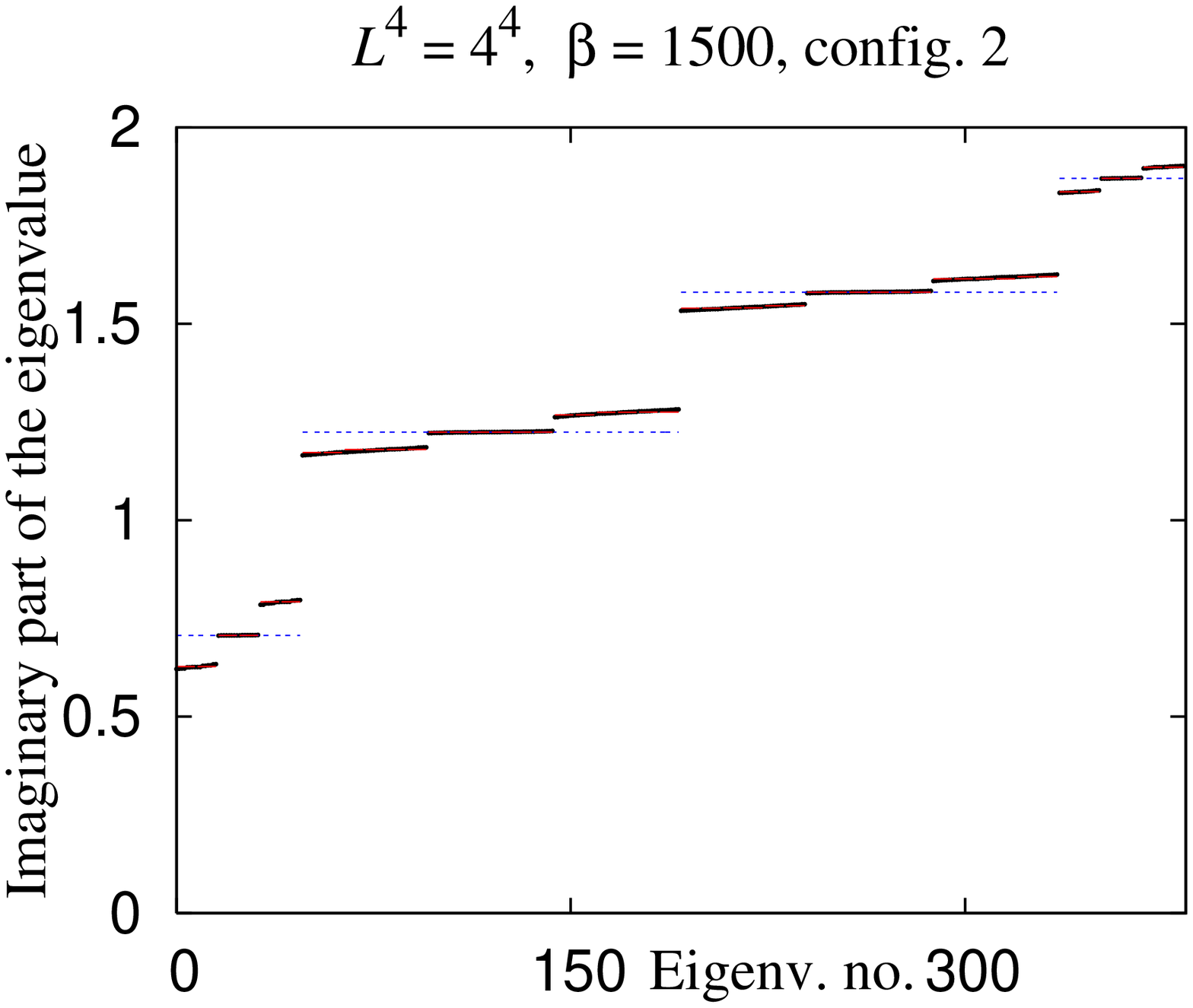}
  \caption{(Color online) Eigenvalues of $\dks$ close to the free limit for the
    adjoint representation of $\SU(2)$ ($L^4=4^4,\,\beta=1500$). The
    unique plateau structure is marked by dashed blue lines.  Within
    the plateaux the eigenvalues (black dots) arrange themselves in
    clusters of eight, whose locations are predicted by
    Eq.~(\protect\ref{adjointclustersandpolyakovloops}) (solid red
    lines).}
  \label{adjointL4polyakovfig}
\end{figure*}

When $\beta$ is increased to large values, on a lattice with a
fixed number of sites in each of the four directions, the spectrum of
$\dks$ shows structure on three different scales, see
Figs.~\ref{L4polyakovfig} and \ref{adjointL4polyakovfig} for examples
with fundamental and adjoint fermions, respectively.

\subsubsection{Plateaux}

The coarse structure is given by the free limit, i.e., the eigenvalues
approach the values of Eq.~\eqref{trivialeigenvalues}, with $c_\mu \in
\left\{ 0, \frac{1}{2} \right\}$ chosen to reflect the sign of the
Polyakov loops as described below Eq.~\eqref{trivialeigenvalues}.  The
collection of all eigenvalues in the vicinity of the values predicted
by Eq.~\eqref{trivialeigenvalues} we call a \emph{plateau}, indicated
by dashed blue lines in Figs.~\ref{L4polyakovfig} and
\ref{adjointL4polyakovfig}.

As discussed in Sec.~\ref{polyakovfundsubsect}, for the fundamental
representation there are five different plateau structures. We find all
these classes in our simulated configurations and show representatives
for each class in Fig.~\ref{L4polyakovfig}. (The spectra shown are the
first representatives of each class which we come across in the Monte
Carlo history.)

For adjoint fermions the plateau structure is uniquely determined by
the lattice sizes (and the boundary conditions) since the adjoint
representation is center-blind, as discussed in
Sec.~\ref{polyakovadjsubsect}. Thus, configurations with different
signs of $P_\mu$ lead to the same plateau structure, see
Fig.~\ref{adjointL4polyakovfig}.

\subsubsection{Clusters}

A closer look at the staggered spectra reveals that the distribution
of eigenvalues inside a given plateau shows additional structure. The
eigenvalues are grouped in \emph{clusters of eight}. When the free
limit is approached, the typical separation between nearest clusters
within the same plateau is smaller than the separation between
different plateaux, but larger than the separation of eigenvalues
within each cluster.

The position of the clusters can be predicted by
Eqs.~\eqref{clustersandpolyakovloops} and
\eqref{adjointclustersandpolyakovloops} for the fundamental and
adjoint representation, respectively. In Figs.~\ref{L4polyakovfig} and
\ref{adjointL4polyakovfig} these predictions are indicated by solid
red lines.  Thus, the cluster structure is determined by the traces of
the averaged Polyakov loops (and the lattice size and boundary
conditions). For the adjoint representation, we see that one-third of
the eigenvalues forming a plateau does not split into clusters, but
stays close to the plateau levels. This is also in agreement with our
discussion of Eq.~\eqref{adjointclustersandpolyakovloops}.

Recall that, approaching the free limit, the distribution of the
traced Polyakov loops becomes peaked at $\pm 1$, corresponding to the
center elements of $\SU(2)$. This \emph{a posteriori} justifies the
approximation of the staggered spectrum by the plateaux.

The observation that each cluster contains eight eigenvalues reflects
the fact that the eigenvalues of Eqs.~\eqref{clustersandpolyakovloops}
and \eqref{adjointclustersandpolyakovloops} are degenerate with
multiplicity 8. These eigenvalues where calculated for vacuum
configurations with uniform and commuting links $U_\mu(x)$ in each
direction. Since the simulated configurations close to the free limit,
which we approximate by these vacuum configurations, do not have
exactly uniform and commuting links, the eightfold degeneracy is
lifted.

In the lower right panel of Fig.~\ref{L4polyakovfig} we also show an
example obtained from simulations on a $6^4$ lattice. We see that for
larger lattices the expected patterns get more and more complicated,
but the agreement persists with just the four parameters $P_\mu$ of
Eq.~(\ref{clustersandpolyakovloops}) determining the complete cluster
structure.

\begin{figure*}
  \centering
  \includegraphics[width=.32\textwidth]{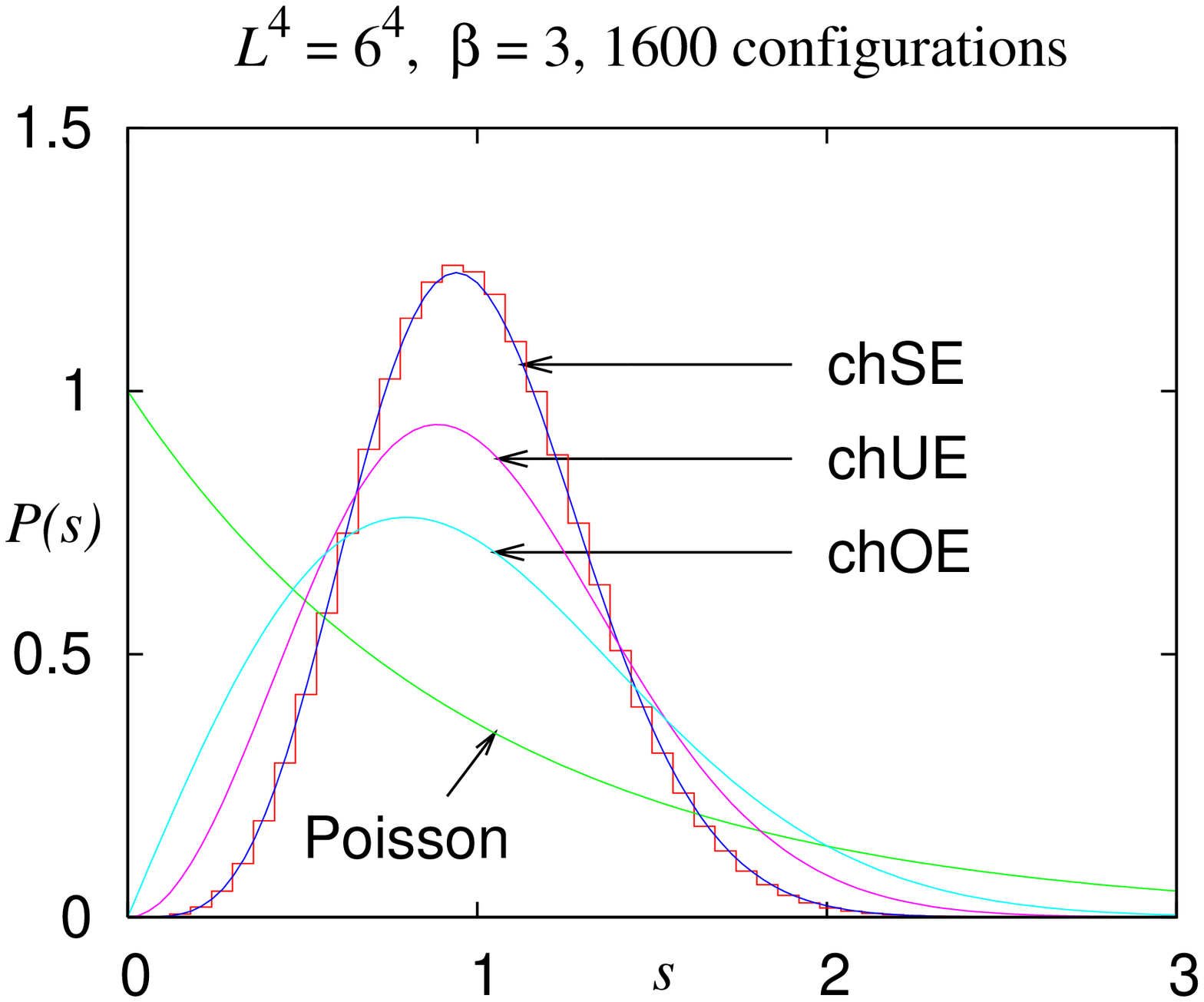}\hfill
  \includegraphics[width=.32\textwidth]{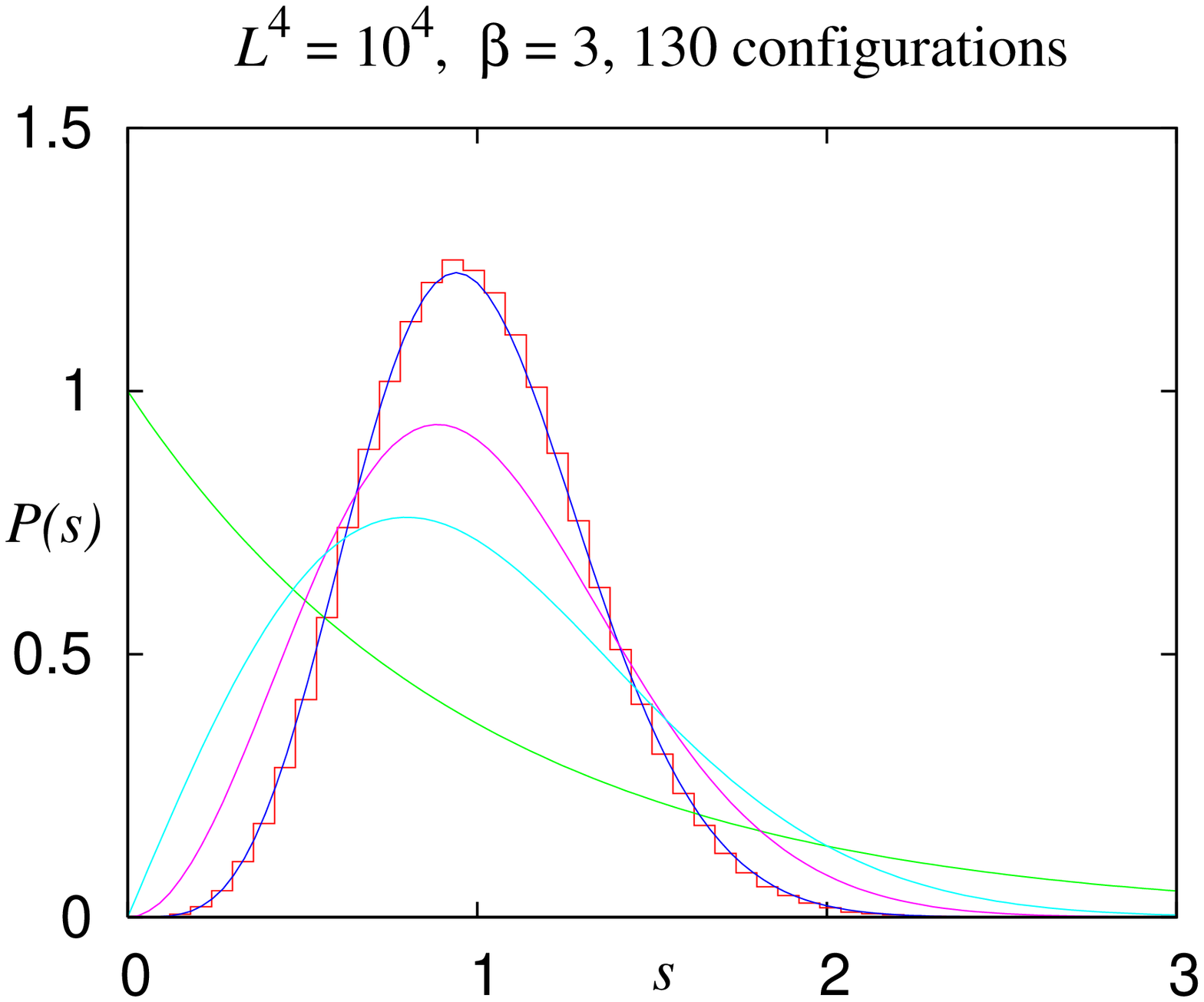}\hfill
  \includegraphics[width=.32\textwidth]{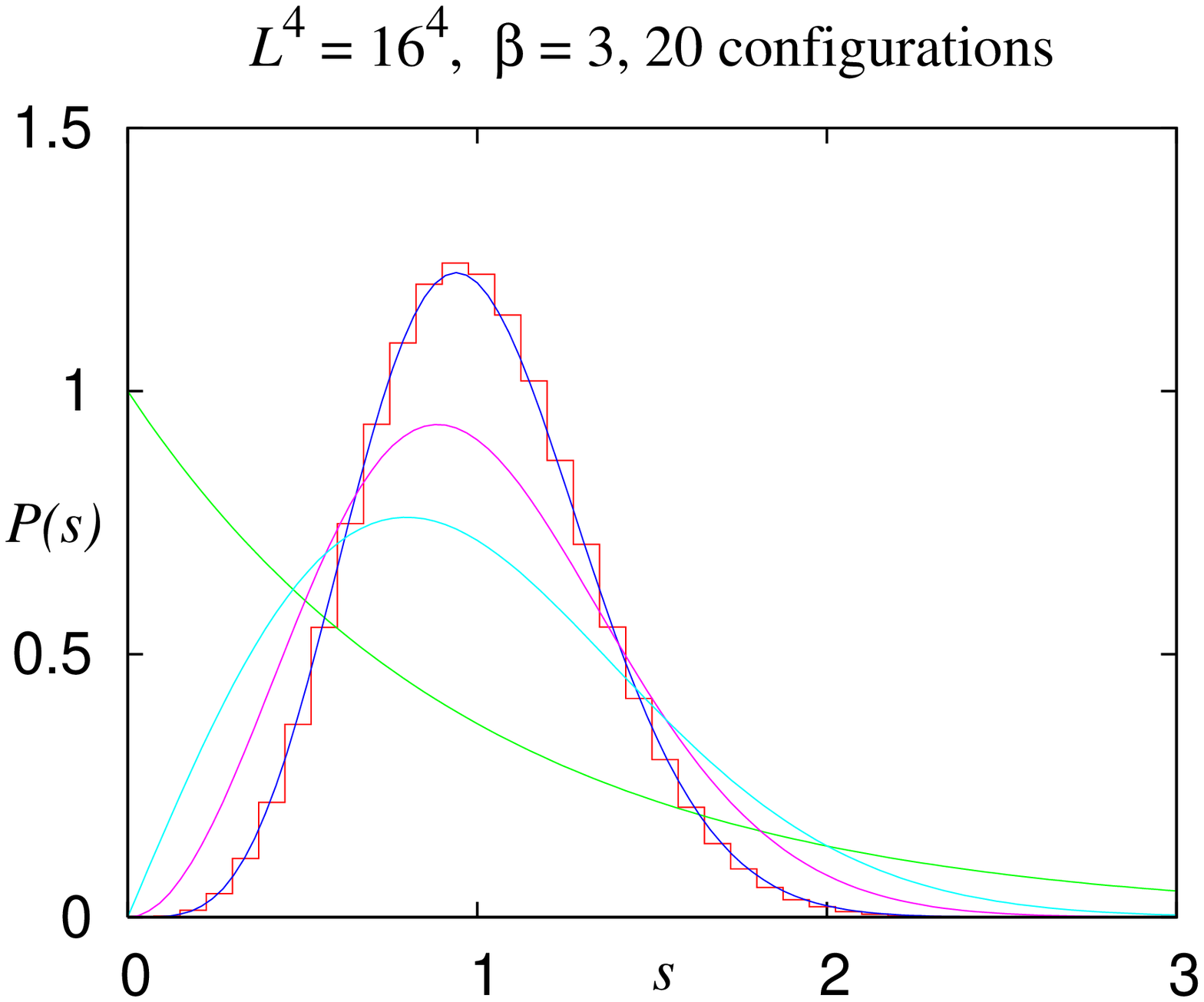}\\
  \caption{(Color online) Level spacing densities obtained from spectra of the
    staggered Dirac operator in the fundamental representation of
    $\SU(2)$ for small $\beta$. The numerical data are consistent with
    the Wigner surmise for the chSE -- as expected according to the
    symmetry properties of the staggered Dirac operator, see
    Sec.~\ref{antiunitaryfundsubsect}}
  \label{smallbetachse}
\end{figure*}

\subsection{Spectral statistics}

\subsubsection{Unfolding}
\label{sec:unfolding}

Before one can discuss spectral statistics and compare to, e.g.,
predictions from RMT, the spectra have to be unfolded, i.e., the
(imaginary parts of the) eigenvalues have to be rescaled such that the
mean separation between adjacent levels is unity. This can be
achieved by defining a new spectrum $\{ x_n \}$ with $x_n :=
\overline{N}(\lambda_n)$, where $\overline{N}(\lambda)$ denotes the
mean integrated spectral density, i.e., $\overline{N}(\lambda)$ is a
smooth function satisfying $\overline{N}(\lambda) \approx
\#\{\lambda_n \leq \lambda\}$ in some approximate or asymptotic sense.

In low-dimensional quantum chaos, one usually has an analytical formula
for $\overline{N}(\lambda)$, provided by Weyl's law, see, e.g.,
\cite{HaakeBook}. For the QCD Dirac operator first steps towards an
equivalent asymptotics have been discussed in \cite{GuhKep07}. Until
now, however, this approach does not provide a formula which can be
used for unfolding spectra in lattice QCD.

If one has no analytical prediction for the mean spectral density, it has to be extracted from the data themselves. The latter can be done by averaging over several spectra, which is known as ensemble
unfolding, thus yielding one mean density for a whole ensemble of
spectra. In contrast, fitting an ansatz to the spectral density of an
individual spectrum or extracting its mean density by a moving average
in $\lambda$ is known as configuration unfolding. The resulting mean
densities will in general differ from each other. This ambiguity can
make it 
difficult to extract reliable information on long-range spectral
correlations, see the detailed discussion of different unfolding
methods for lattice Dirac spectra in Ref.~\cite{Guhr:1998pa}. The
statistical function we are interested in is always the density $P(s)$
of nearest-neighbor spacings. Since $P(s)$ measures short-range
spectral correlations, changing the unfolding method will not impair our results, as long as we make sure that each
unfolding procedure is stable and consistent.

As we want to discuss spectral statistics for very different values
of the gauge action parameter $\beta$ and on different scales, we are
forced to employ different unfolding methods, each one of them
tailored to fit the requirements of the particular situation. 

For large $\beta$ the spectra show different scales, see
Sec.~\ref{separationofscalessec}. Therefore, we unfold separately on
each scale. The different plateau and cluster structures forbid
methods of ensemble unfolding, and thus we employ configuration
unfolding. The methods we use are all variants of what is called local
unfolding in Ref.~\cite{Guhr:1998pa}.

When studying spacings within clusters we divide spacings of adjacent
levels by the mean level spacing within the cluster. The latter we
calculate as the difference between the largest and the smallest level
divided by seven. The unfolded spacings then have unit mean as
required.

For spectral statistics between clusters within a plateau, we proceed
analogously. We divide spacings between adjacent clusters by the
difference between the position of the largest and the smallest
cluster divided by the number of clusters within the plateau minus
one.

Finally, when we want to study spacings between plateaux, we adapt the
previous methods as follows. We divide the difference between the
position of plateau $j+1$ and plateau $j$ by the difference between
the positions of plateaux $j+5$ and $j-4$ divided by the number of
plateau spacings in this range, i.e., by nine.

For small $\beta$ the spectra do not show different scales. Thus, complete spectra could be unfolded in one go, and both configuration
and ensemble unfolding are admissible. We experimented with the
ensemble unfolding described in Ref.~\cite{Edwards:1999ra}, which
works reliably for small $\beta$. However, when we increase $\beta$
the method has to break down, due to the different plateau structures
which begin to emerge. In order to have an unfolding method which does
not exclude an intermediate $\beta$-range but instead allows us to go
smoothly from small to large $\beta$, we decided to unfold spectra for
small $\beta$ in the same way as we unfold spacings within clusters
for large $\beta$. Effectively, this means that we neglect every
eighth spacing and unfold the remaining spacings on the scale of the
neighboring eight eigenvalues.

\subsubsection{Level spacings for small $\beta$}
\label{smallbetasubsect}

For small values of $\beta$, i.e., far away from the free and
continuum limits, the level spacing densities have to be compared to
the predictions reported in Secs.~\ref{antiunitaryfundlatsubsubsect}
and \ref{antiunitaryadjlatsubsubsect}.

Figure \ref{smallbetachse} shows the level spacing densities obtained
from spectra of the staggered Dirac operator in the fundamental
representation of $\SU(2)$. The numerical results on all our lattices
are consistent with the chSE, in contrast to the chOE, which would be
expected in the continuum.

\begin{figure*}
  \centering
  \includegraphics[width=.32\textwidth]{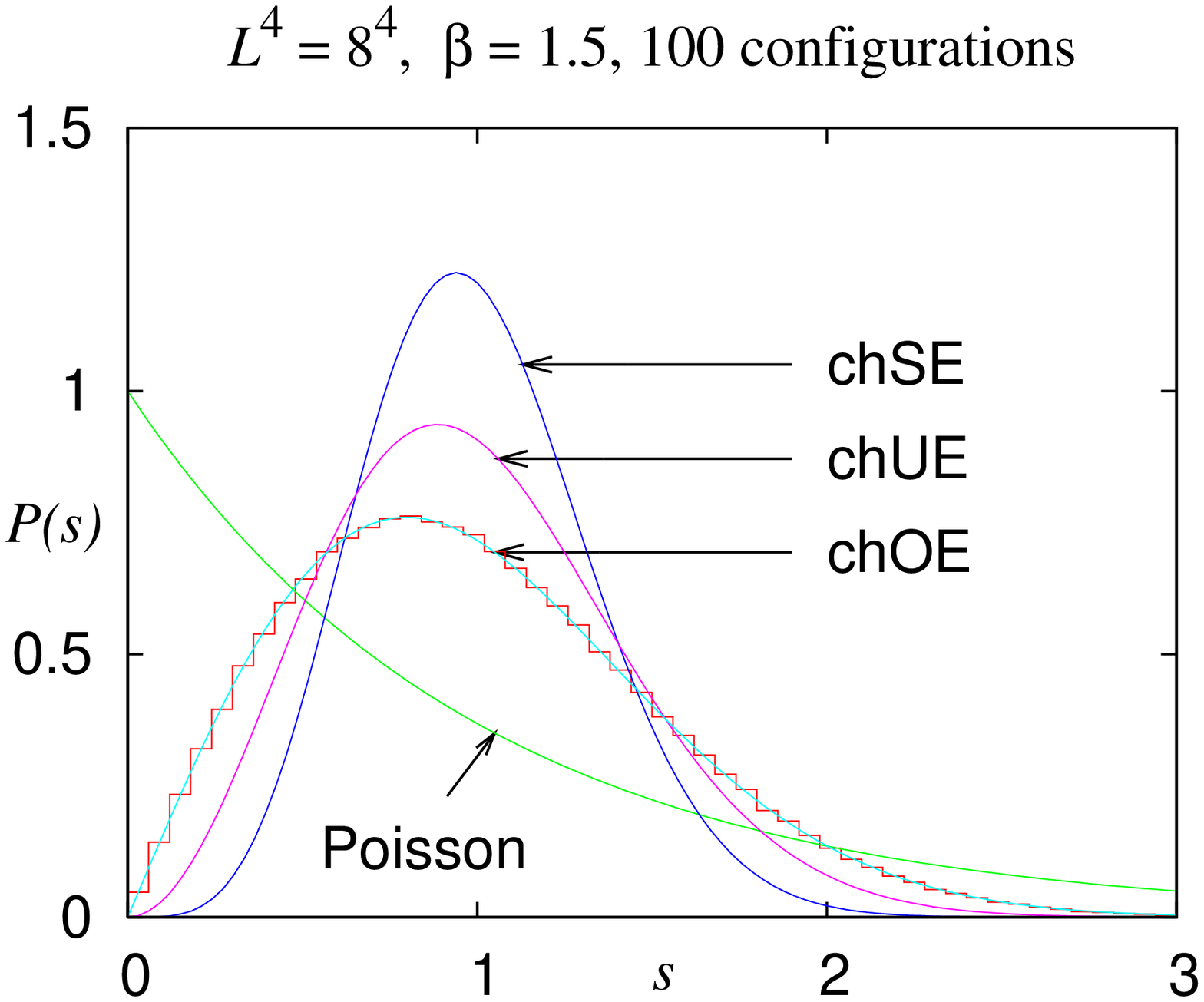}
  \includegraphics[width=.32\textwidth]{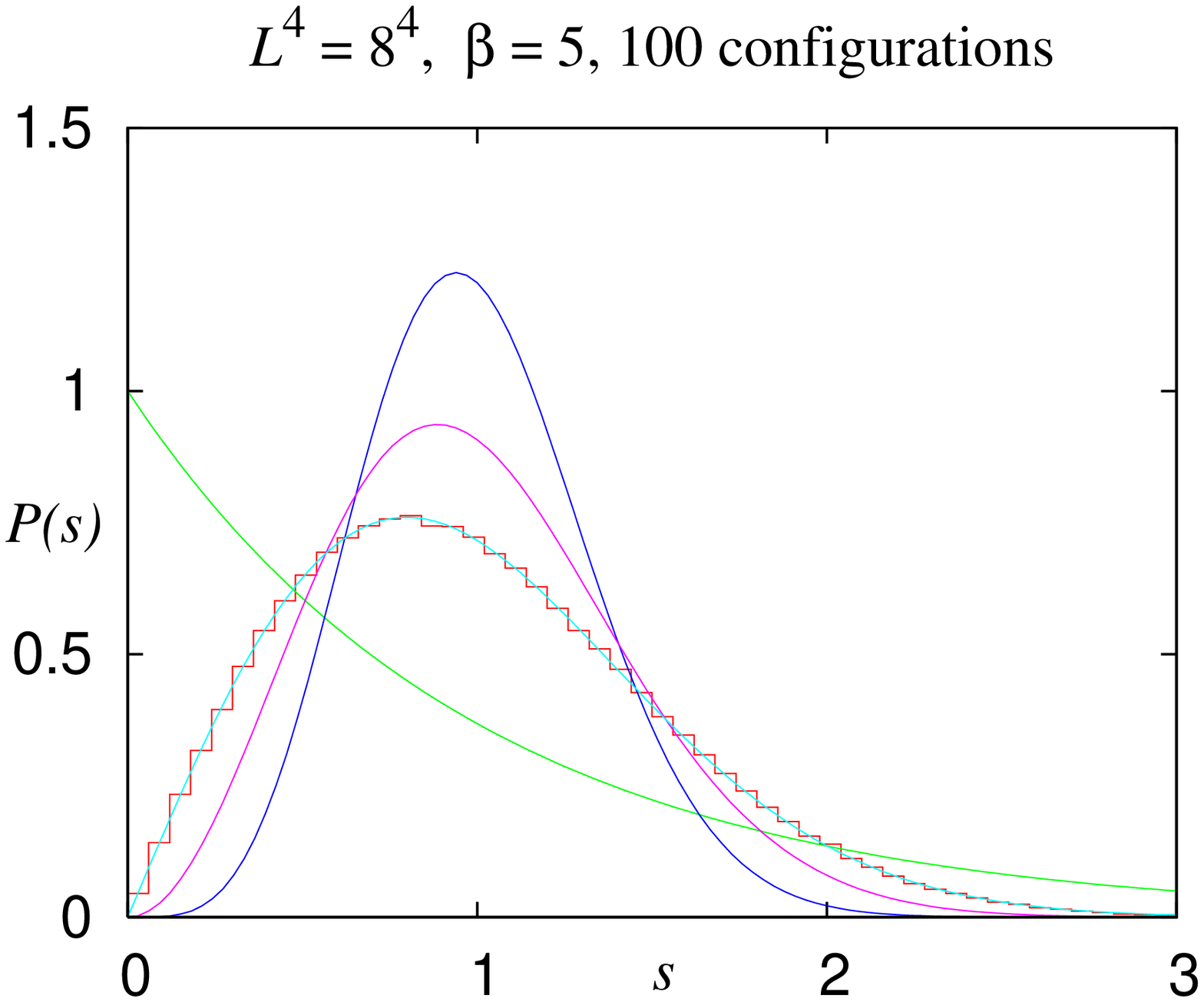}
  \caption{(Color online) Level spacing densities obtained from spectra of the
    staggered Dirac operator in the adjoint representation of $\SU(2)$
    for small $\beta$. The numerical data are consistent with the
    Wigner surmise for the chOE -- as expected according to the
    symmetry properties of the staggered Dirac operator, see
    Sec.~\ref{antiunitaryadjsubsect}.}
  \label{smallbetachoe}
\end{figure*}

Likewise, in Fig.~\ref{smallbetachoe} we show level spacing densities
obtained from staggered Dirac spectra in the adjoint representation of
$\SU(2)$, which now are consistent with the chOE as expected. Again this is in contrast to the continuum situation in which one should have
chSE statistics.

This behavior of the staggered Dirac operator in $\SU(2)$ gauge
fields, having spectra in different universality classes than the
corresponding continuum operator, has been observed earlier
\cite{Halasz:1995vd,BerbenniBitsch:1997tx,Edwards:1999px}.

\begin{figure*}
  \centering
  \includegraphics[width=.32\textwidth]{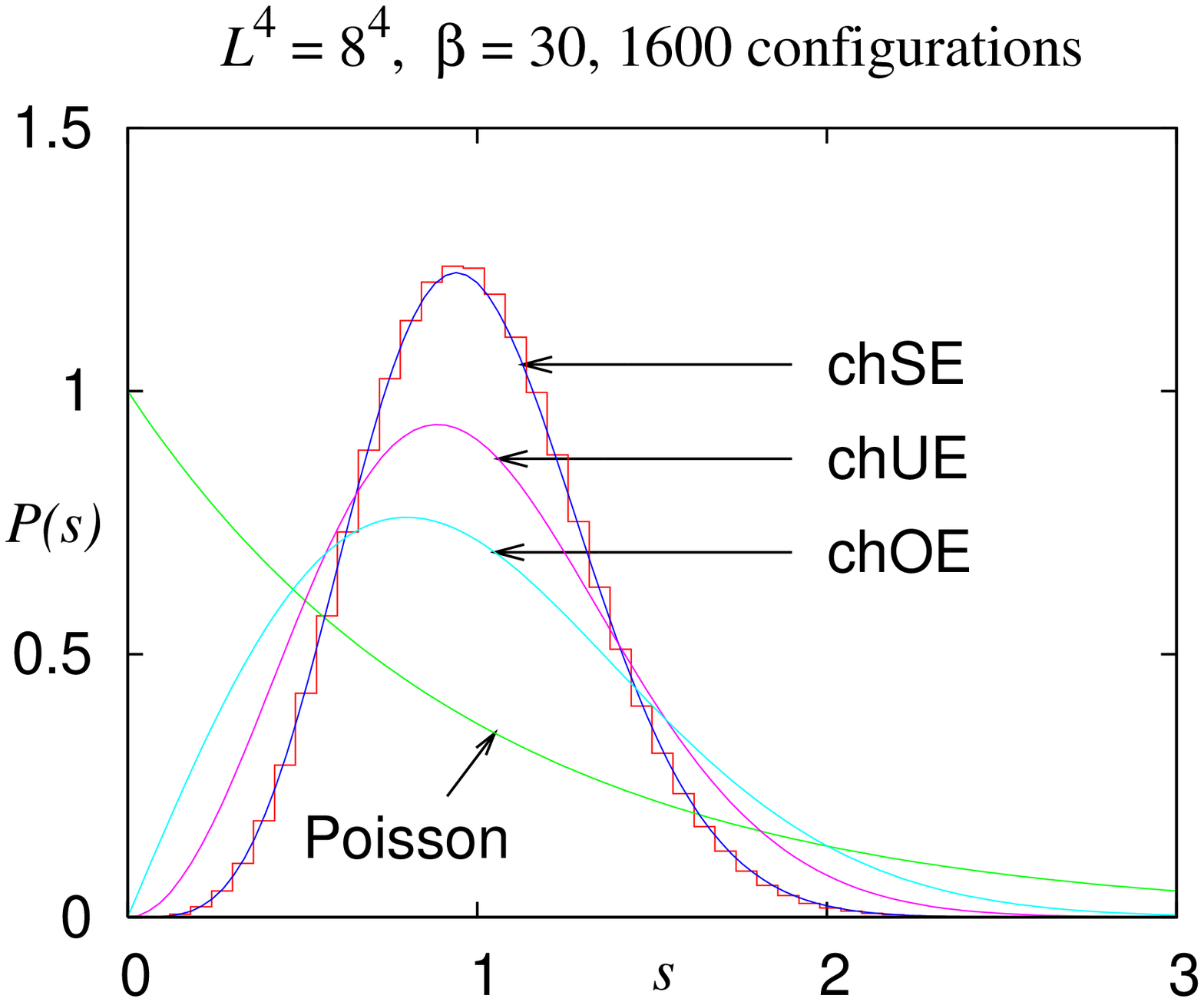}\hfill
  \includegraphics[width=.32\textwidth]{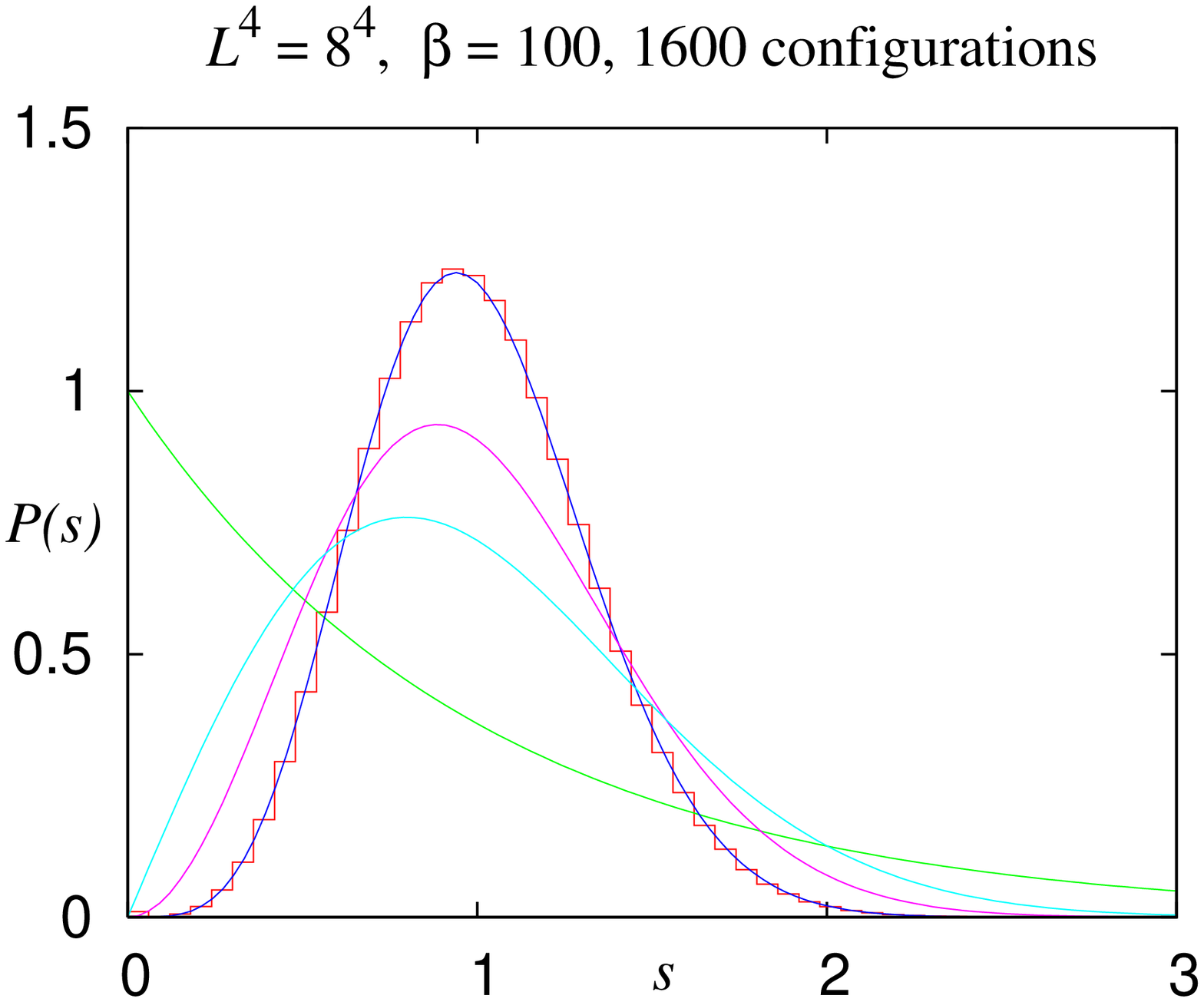}\hfill
  \includegraphics[width=.32\textwidth]{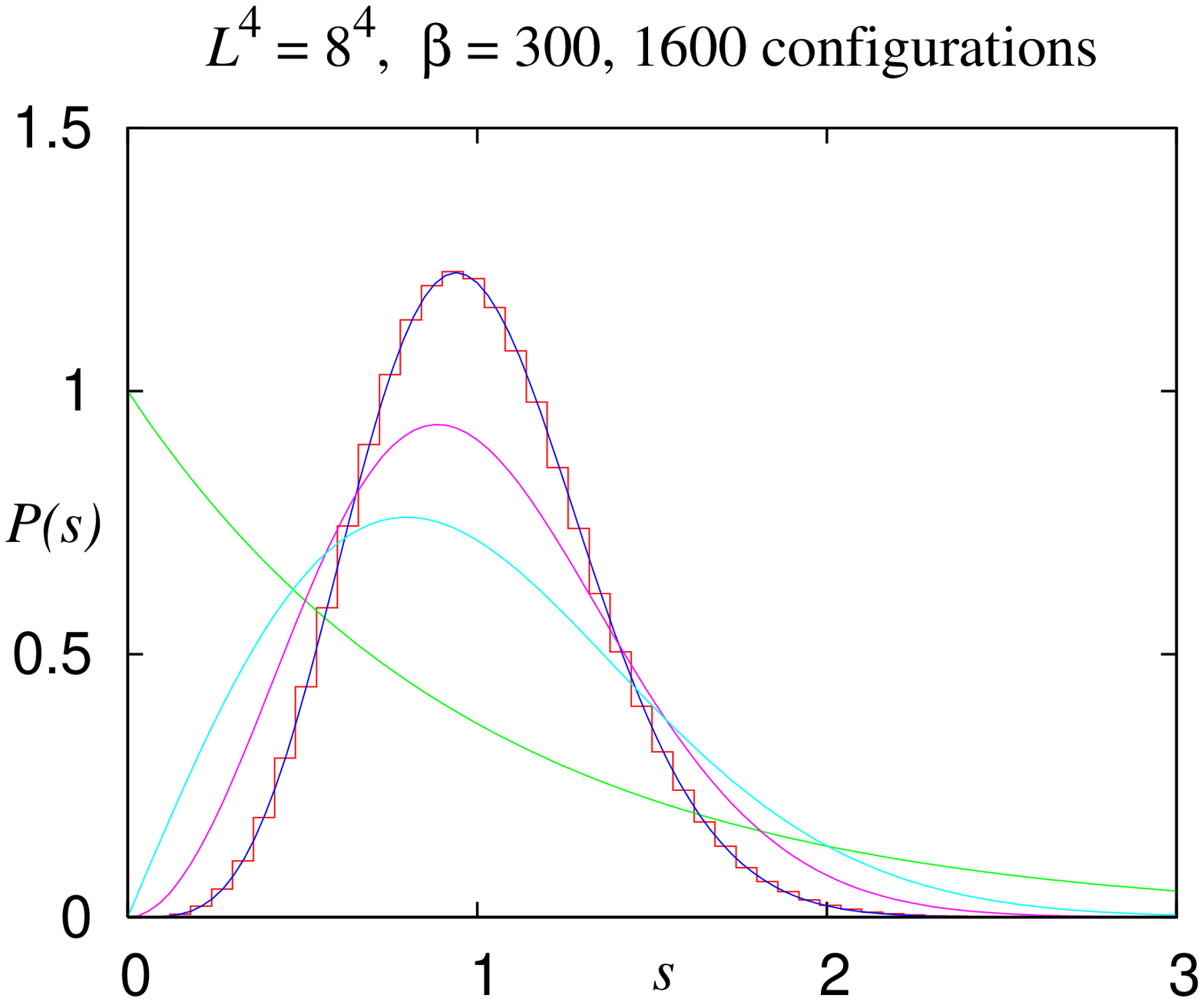}\\
  \includegraphics[width=.32\textwidth]{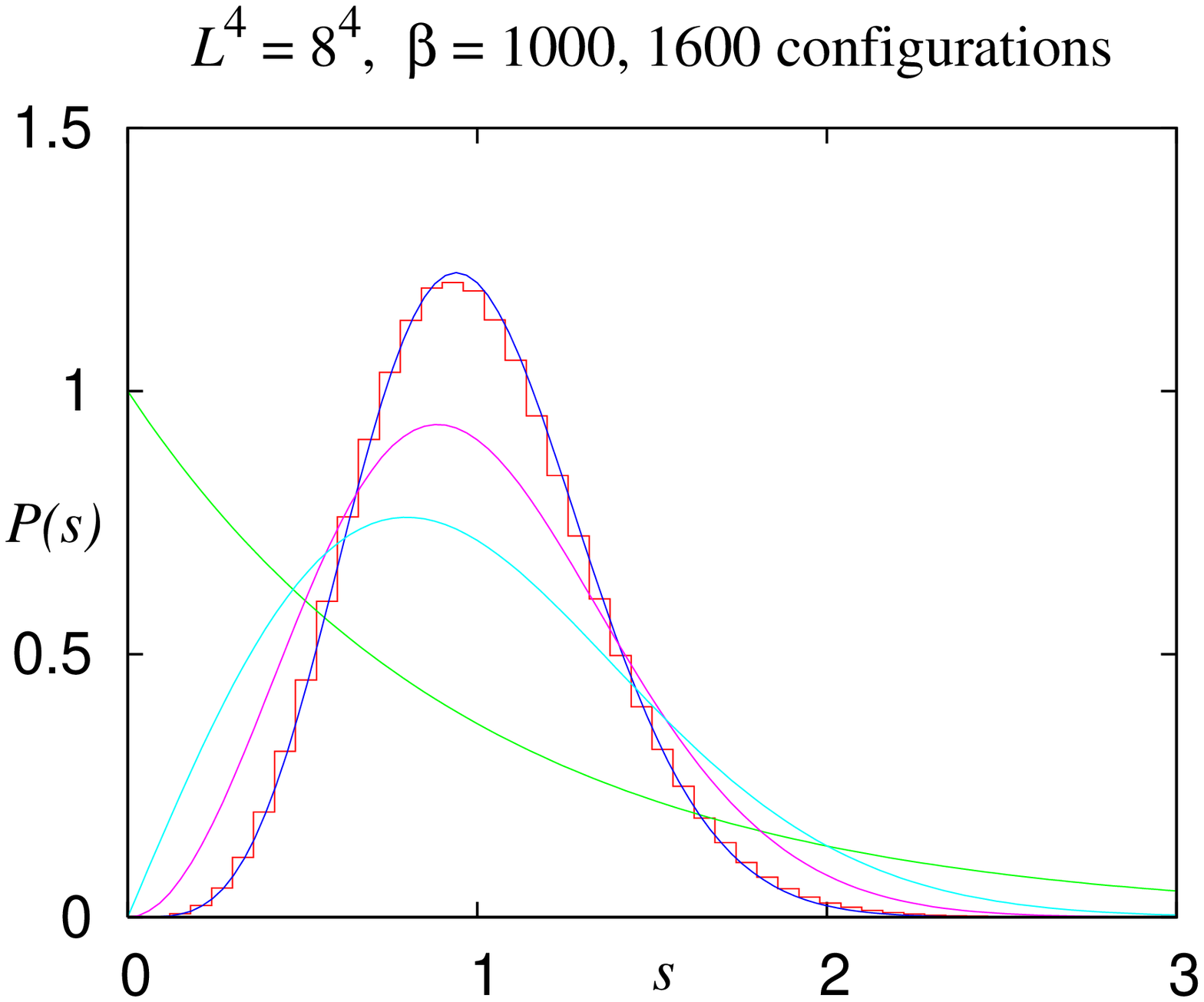}\hfill
  \includegraphics[width=.32\textwidth]{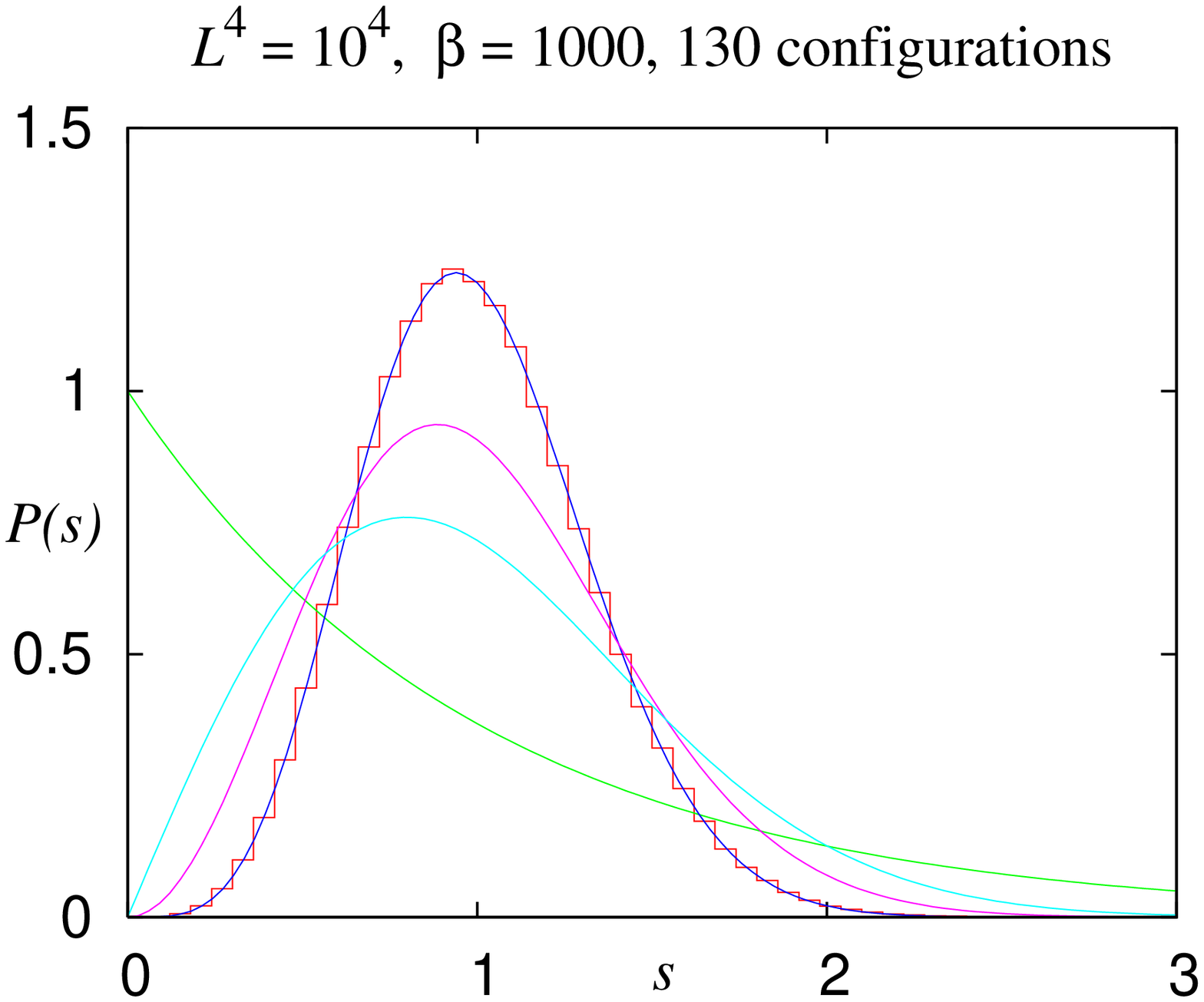}\hfill
  \includegraphics[width=.32\textwidth]{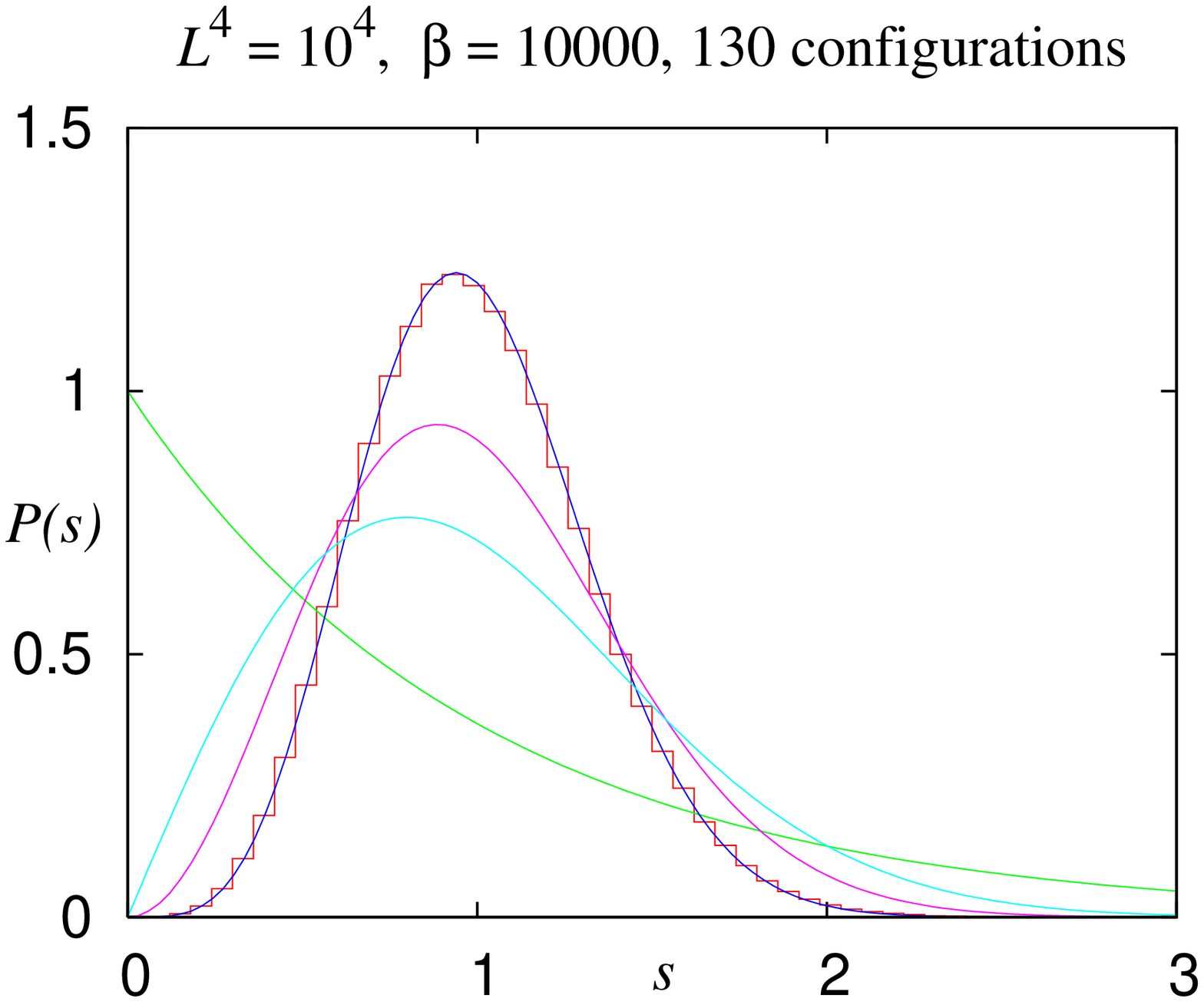}
  \caption{(Color online) Level spacing densities for eigenvalues within clusters,
    obtained from spectra of the staggered Dirac operator in the
    fundamental representation of $\SU(2)$ for increasing $\beta$.
    Agreement with the chSE persists for large $\beta$, i.e., close to
    the free limit.}
  \label{stillchse}
\end{figure*}

\begin{figure*}
  \centering
  \includegraphics[width=.32\textwidth]{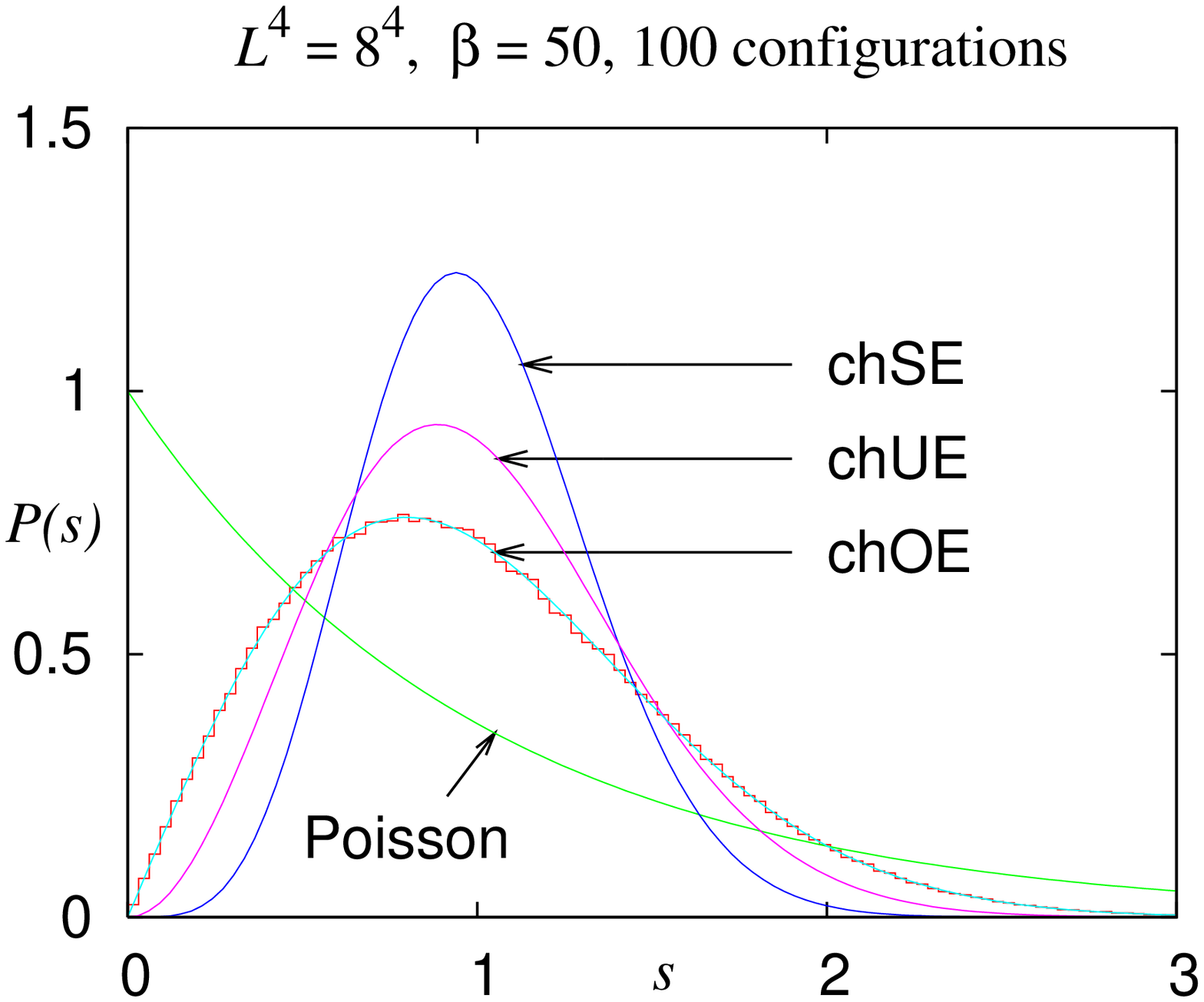}\hfill
  \includegraphics[width=.32\textwidth]{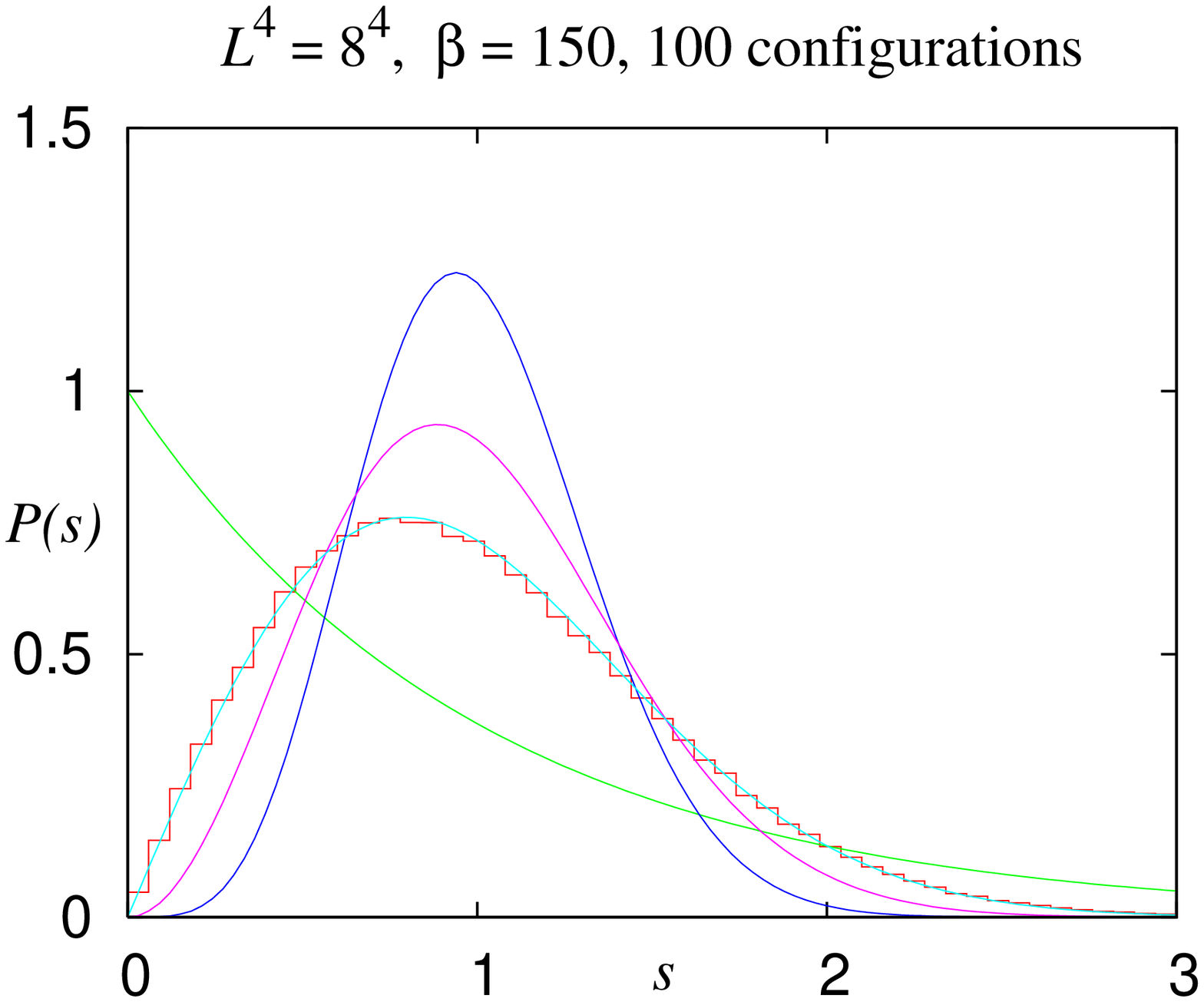}\hfill
  \includegraphics[width=.32\textwidth]{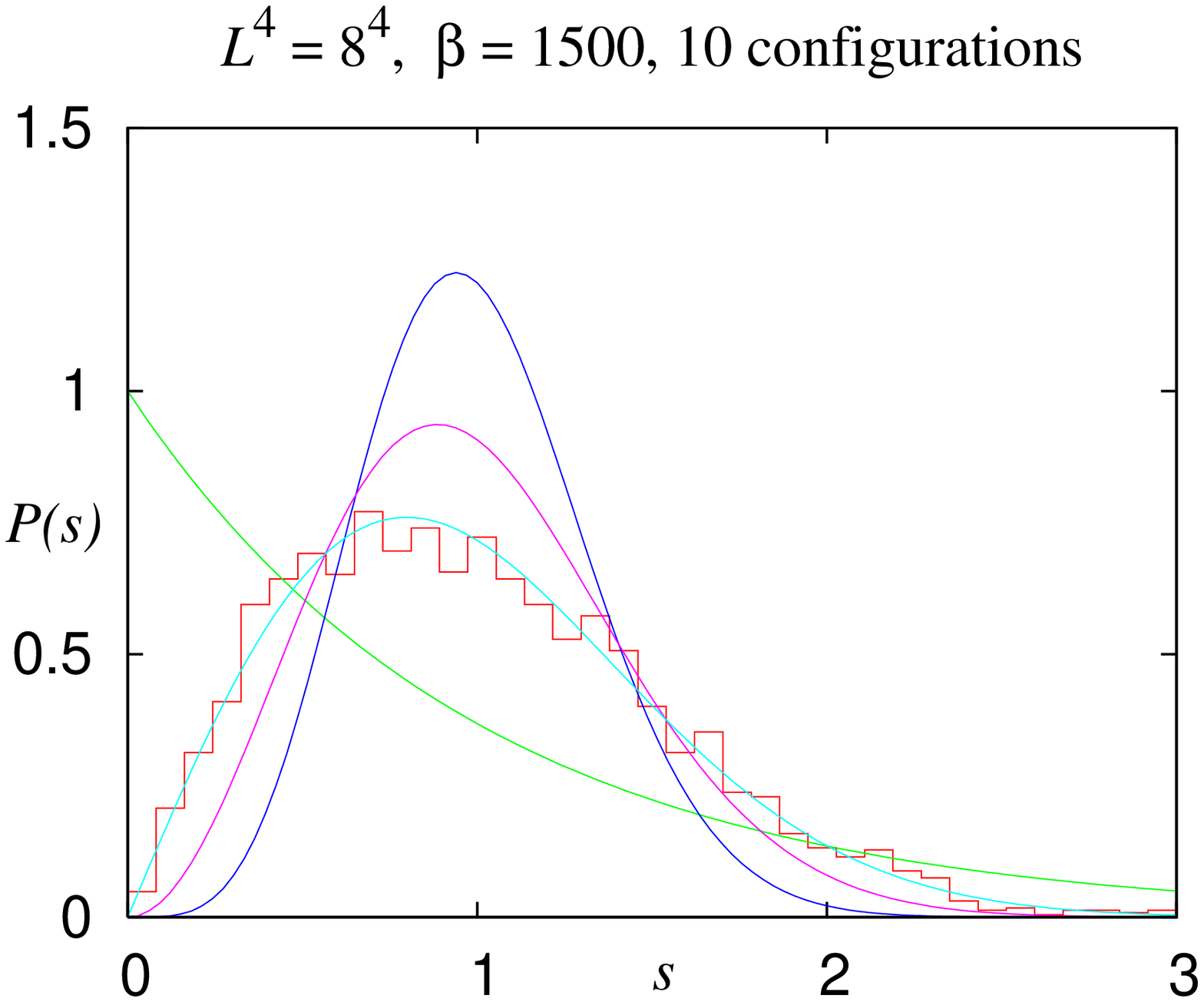}
  \caption{(Color online) Level spacing densities for eigenvalues within clusters,
    obtained from spectra of the staggered Dirac operator in the
    adjoint representation of $\SU(2)$ for increasing $\beta$. In the
    central part of each plateau (cf.\ text) agreement with the chOE
    persists for large $\beta$, i.e., close to the free limit.}
  \label{freelimitchoe}
\end{figure*}

\subsubsection{Level spacings for large $\beta$: Approaching the free limit}

For large $\beta$, plateaux and clusters emerge. Therefore, we discuss
spectral statistics separately on the different scales.

(a) {\it Spacings within clusters.} Generically, the eightfold
degeneracy of the levels predicted by
Eqs.~\eqref{clustersandpolyakovloops} and
\eqref{adjointclustersandpolyakovloops} is lifted by the
nonuniformity of the gauge field configuration. For this reason we
observe small clusters of eight eigenvalues.

Close to the free limit we can think of treating the nonuniformity of
the gauge field configuration as a small perturbation of the
corresponding vacuum configuration, i.e., of the configuration with
uniform and commuting links leading to the same Polyakov loops. A
cluster then arises by diagonalizing an $8 \times 8$ matrix, the
perturbation restricted to the subspace corresponding to a degenerate
eigenvalue of Eq.~\eqref{clustersandpolyakovloops} or
Eq.~\eqref{adjointclustersandpolyakovloops}. This matrix inherits
symmetries and effective randomness of the perturbation, i.e., of the
gauge field part of the staggered Dirac operator. Therefore, we expect
the distribution of level spacings to follow the same chRMT prediction
as for small $\beta$.

Our data for fermions in the fundamental representation of $\SU(2)$
confirm this expectation. As Fig.~\ref{stillchse} shows, the level
spacing density within each cluster is consistent with the prediction
from the chSE. Note that this is true over a very large range of
$\beta$ values.

Also in the adjoint representation the spacings within clusters follow
the same pattern as the spacings for small $\beta$, in this case
leading to a chOE distribution, see Fig.~\ref{freelimitchoe}.  Note
that for the adjoint representation, the clusters at the ends of a
plateau show some nongeneric features which would require a more
sophisticated unfolding procedure.  We avoid this complication by
restricting the analysis to cluster spacings in the central part of
each plateau.

These results confirm once more that the spectral properties of the
staggered Dirac operator are different from those of the continuum
operator. Moreover, in the sense described above, our analysis
demonstrates that this discrepancy persists when approaching the free
limit.

(b) {\it Spacings between clusters.} For the fundamental
representation the cluster positions are, to a good approximation,
described by Eq.~\eqref{clustersandpolyakovloops}. Both
Eq.~\eqref{trivialeigenvalues} and Eq.~\eqref{clustersandpolyakovloops}
describe spectra of integrable systems, and thus one generically
expects uncorrelated levels, like for a Poisson process
\cite{BerTab77,Bohigas:1983er}.

However, on commensurate lattices, i.e., on lattices with rationally
dependent $L_\mu/2$, Eq.~\eqref{trivialeigenvalues} predicts many
accidental degeneracies which would lead to nongeneric spectral
statistics. Generic behavior can in principle be restored in two
different ways: On the one hand by changing to lattices with
rationally independent extensions $L_\mu/2$, and on the other hand by
introducing additional phase shifts, like the Polyakov loops do in
Eq.~\eqref{clustersandpolyakovloops}. However, as we pointed out
earlier, close to the free limit the distribution of averaged traced
Polyakov loops is peaked at $\pm 1$. In Fig.~3 of
Ref.~\cite{Bruckmann:2008rj} it has been observed that these almost
equal phase shifts in each direction are not able to restore
generic behavior.

\begin{figure*}
  \includegraphics[width=.32\textwidth]{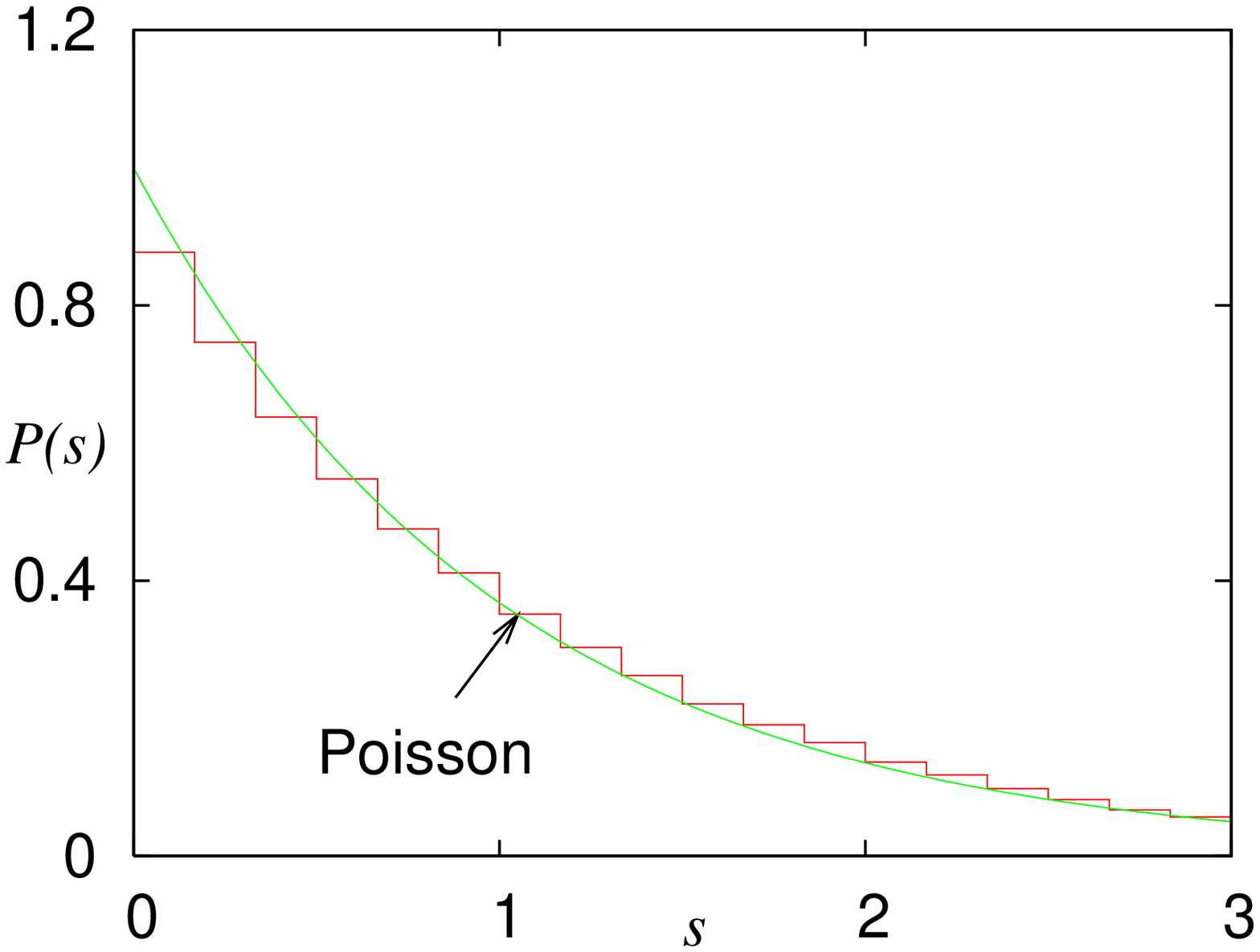}\hfill
  \includegraphics[width=.32\textwidth]{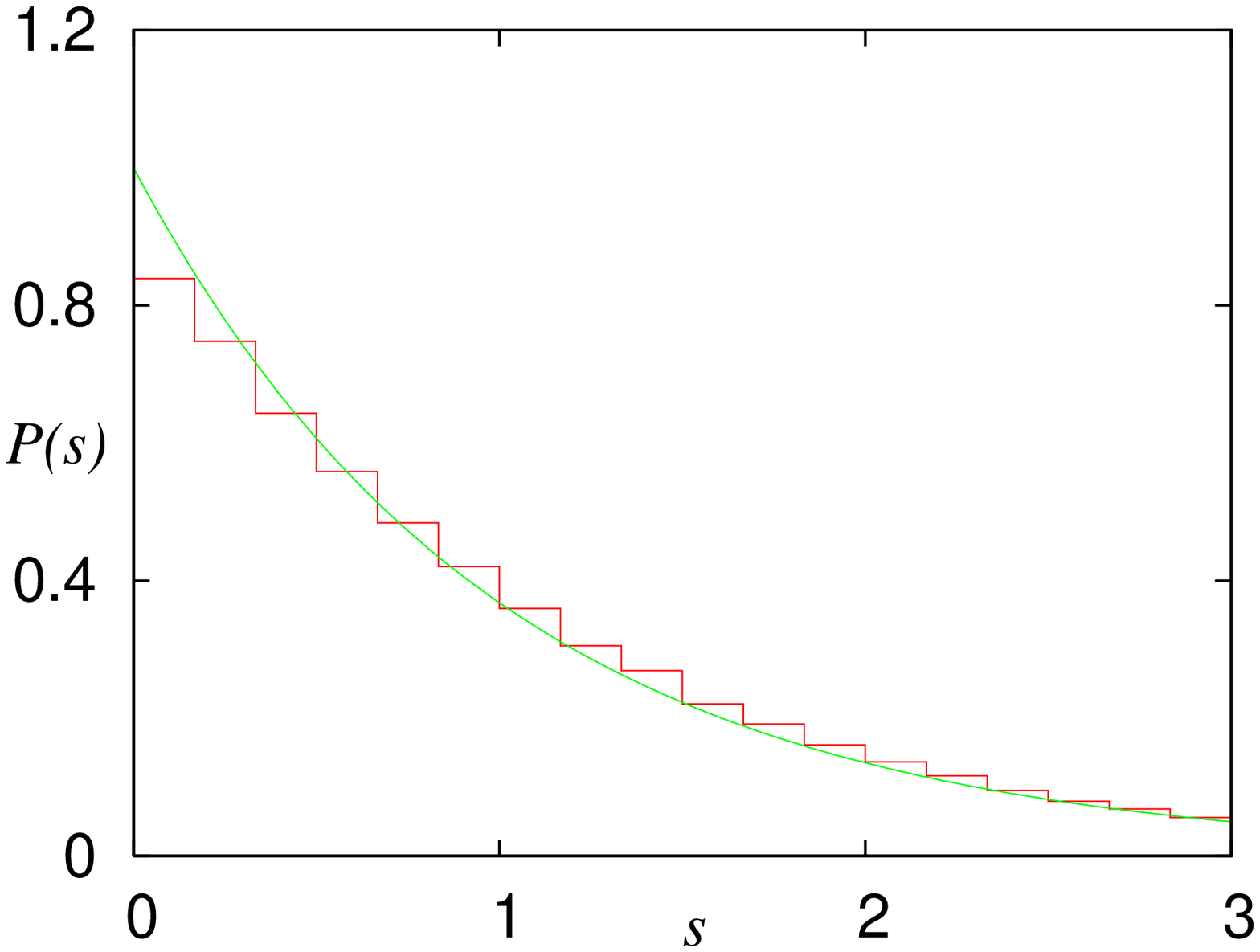}\hfill
  \includegraphics[width=.32\textwidth]{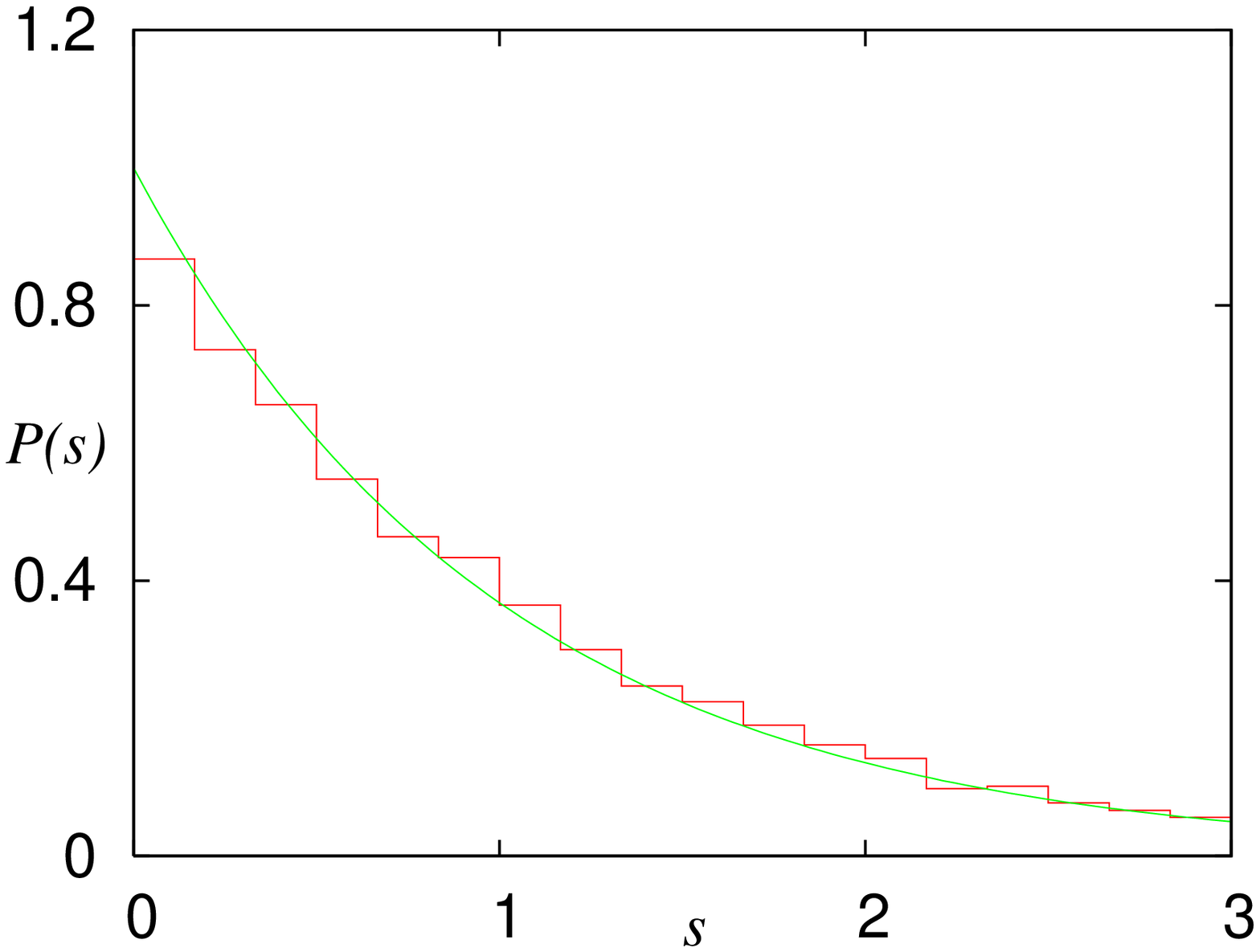}
  \caption{(Color online) Left: The spacing density of the approximate cluster
    spectrum of the fundamental representation predicted by
    Eq.~\protect\eqref{clustersandpolyakovloops} for a single
    configuration on a $34 \times 38 \times 46 \times 58$ lattice at
    $\beta=10000$ agrees with the spacing density $e^{-s}$ for a
    Poisson process.  Middle: Same as left panel, but now for the
    adjoint representation and
    Eq.~\protect\eqref{adjointclustersandpolyakovloops}.  Right: The
    spacing density of the plateau positions, i.e., the eigenvalues of
    the free staggered Dirac operator, as predicted by
    Eq.~\eqref{trivialeigenvalues} on a $34 \times 38 \times 46 \times
    58$ lattice also agrees with the spacing density $e^{-s}$ for a
    Poisson process.}
  \label{poissonlargelatticefig}
\end{figure*}

Therefore, we choose to study a large incommensurate lattice with
extensions $L_1 \times L_2 \times L_3 \times L_4 = 34 \times 38 \times
46 \times 58$. We refrain from diagonalizing the Dirac operator on
this lattice but instead calculate the averaged traced Polyakov loops
and determine the approximate cluster spectrum from
Eq.~\eqref{clustersandpolyakovloops}. Then the density of spacings
between different clusters within the same plateau agrees well with
the prediction from a Poisson process, $P_\mathrm{Poisson}(s) =
e^{-s}$, see Fig.~\ref{poissonlargelatticefig} (left).  The same holds
true for fermions in the adjoint representation with the approximate
cluster spectrum determined from
Eq.~\eqref{adjointclustersandpolyakovloops}, see
Fig.~\ref{poissonlargelatticefig} (middle).

Note that we analyze the spacing distribution for the spectra
\eqref{clustersandpolyakovloops} and
\eqref{adjointclustersandpolyakovloops} but not the spacings between
the actual cluster positions which one would obtain by averaging over
all eigenvalues belonging to a given cluster. As can be seen in
Figs.~\ref{L4polyakovfig} and \ref{adjointL4polyakovfig} these actual
positions differ slightly from the predicted values. Whether or not
these differences could lead to a deviation from Poisson behavior is
an open question.

(c) {\it Spacings between plateaux.}  The positions of the plateaux
can be approximately described by the free spectrum,
Eq.~\eqref{trivialeigenvalues}.  As in the case of spacings between
clusters, which we discussed above, we thus once more study the
spectrum of an integrable system. Again accidental degeneracies lead
to nongeneric statistics on commensurate lattices. Generic behavior
can be restored by switching to an incommensurate lattice, which we
demonstrate in Fig.~\ref{poissonlargelatticefig} (right). Level
spacing distributions like for a Poisson process were predicted and
observed earlier for incommensurate lattices
\cite{Verbaarschot:1997bf,Berg:1999va}.

We add the same note of caution as for the spacings between clusters.
Our analysis concerns the level spacings of the spectrum
\eqref{trivialeigenvalues} but not the spacings between the actual
plateau positions, obtained by averaging over all eigenvalues within a
given plateau.

Let us also point out that, in spite of the strong similarities
between the data describing the cluster spacings and the plateau
spacings in Fig.~\ref{poissonlargelatticefig}, they describe
correlations on spectral scales that typically differ by an order of
magnitude.

\section{Conclusions and outlook}
\label{conclusionssect}

We have studied the spectral properties of the staggered Dirac
operator $\dks$ when approaching the free limit for gauge group
$\SU(2)$, both in the fundamental and in the adjoint representation.
With $\SU(2)$ gauge fields the staggered Dirac operator on the lattice
belongs to a different symmetry class than the corresponding continuum
operator.

Our numerical analysis revealed that, when the free limit is
approached, the spectrum of the staggered Dirac operator shows
structure on three well-separated scales. The behavior on the two
coarser scales is characterized by the formation of plateaux and
clusters.

We have shown that for a given gauge field configuration the positions
of plateaux and clusters can be predicted analytically. To this end we
have constructed a vacuum configuration with uniform and commuting
links in each direction, chosen such that they reproduce the averaged
traced Polyakov loops of the original configuration. The Dirac
spectrum in this vacuum configuration yields a good approximation to
the plateau and cluster positions. In turn, plateaux and clusters are
determined by lattice geometry, boundary conditions, and Polyakov
loops alone.

Our model also predicts a systematic degeneracy of the staggered
spectra in nontrivial vacuum configurations: All eigenvalues have a
multiplicity of eight (in addition to Kramers' degeneracy in the
fundamental representation). For typical gauge field configurations
this degeneracy is lifted leading to the formation of clusters of
eight eigenvalues, which we observe numerically.

We have analyzed spectral correlations on all three scales in terms of
the distribution of spacings between adjacent eigenvalues, clusters,
and plateaux. Spacings between approximate plateau and cluster
positions are uncorrelated as for a Poisson process, whereas level
spacings on the finest scale, i.e., within clusters, follow the chRMT
predictions. For the latter the symmetry class is always that of the
staggered operator and never that of the continuum operator, even for
very large $\beta$, i.e., close to the free limit.

Finally, we briefly comment on the possibility to recover the symmetry
class of the continuum operator in spectra of the staggered Dirac
operator in $\SU(2)$ gauge fields. We restrict ourselves to the
fundamental representation.  For every finite lattice spacing, the
antiunitary symmetries of the $\dks$ operator are different from
those of the continuum Dirac operator.  When approaching the continuum
limit, i.e., when increasing $\beta$ at fixed physical volume, the
eigenvalues should form near-degenerate quartets (more precisely,
near-degenerate pairs of exactly Kramers-degenerate pairs), signaling
the suppression of unphysical taste-changing interactions of the
staggered operator.  This has been seen for the first time in
Ref.~\cite{Follana:2006zz} using highly improved staggered fermions.
In~\cite{Follana:2006zz} it was also shown that the distribution of
the smallest eigenvalue makes a transition from chSE to chOE, i.e.,
from the symmetry class of the lattice operator to that of the
continuum operator. Related phenomena have been observed for improved
staggered fermions in the fundamental representation of $\SU(3)$
\cite{Follana:2004sz,Durr:2004as,Follana:2005km}. We expect that the
distribution of spacings between eigenvalues will also show a similar
transition to the continuum behavior. Our present findings do
not contradict such an expectation as our study corresponds to a
different physical setting: Instead of the continuum limit we have
investigated the behavior in the free limit, where RMT applies only at
the finest scale of eigenvalue fluctuations.

\begin{acknowledgments}

  We thank T.~Guhr, S.~Schierenberg, and J.J.M.~Verbaarschot for
  helpful discussions and gratefully acknowledge support from the
  Deutsche Forschungsgemeinschaft (F.B., S.K., T.W.) and from the Alexander von Humboldt Foundation (M.P.).

\end{acknowledgments}

\bibliography{references}

\end{document}